\documentclass[aps,prl,twocolumn,showkeys,superscriptaddress,sectionbib,square]{revtex4-1}
\setcitestyle{numbers,square}
\usepackage{epsfig}
\usepackage{color}
\usepackage{times}
\usepackage{amsmath}
\usepackage{amssymb}
\usepackage{indentfirst}
\usepackage{graphicx}
\usepackage{bm}
\usepackage{comment}
\usepackage{wrapfig}

\newcommand{\bey}{\begin{eqnarray}}
\newcommand{\eey}{\end{eqnarray}}
\newcommand{\vep}{\varepsilon}

\newcommand{\sg}{\sigma}
\newcommand{\bec}{\begin{center}}
\newcommand{\eec}{\end{center}}

\begin{document}
\title{In situ quantitative  study of plastic strain-induced phase transformations under high pressure: Example for ultra-pure Zr }

\author{K. K. Pandey}\email{kkpandey@iastate.edu}
\affiliation{Department of Aerospace Engineering, Iowa State University, Ames, Iowa 50011, USA}

\author{Valery I. Levitas}\email{vlevitas@iastate.edu}

\affiliation{Department of Aerospace Engineering, Iowa State University, Ames, Iowa 50011, USA}
\affiliation{Department of Mechanical Engineering, Iowa State University, Ames, Iowa 50011, USA}
\affiliation{Ames Laboratory, U.S. Department of Energy, Iowa State University, Ames, Iowa 50011-3020, USA}

\date{\today}

\begin{abstract}
The first in situ quantitative synchrotron X-ray diffraction (XRD) study of plastic strain-induced  phase transformation (PT) has been  performed on $\alpha-\omega$ PT in ultra-pure, {\color{black} strongly plastically predeformed } Zr as an example,  under different compression-shear pathways in rotational diamond anvil cell (RDAC). Radial distributions of pressure in each phase and in the mixture, and concentration of $\omega$-Zr, all averaged over the sample thickness, as well as thickness profile were measured. The minimum pressure for the strain-induced $\alpha-\omega$ PT, $p^d_{\vep}$=1.2 GPa, is smaller than under hydrostatic loading by a factor of 4.5  {\color{black} and smaller than the phase equilibrium pressure by a factor of 3; it is  independent of the compression-shear straining path}. The theoretically predicted plastic strain-controlled kinetic equation  was verified and quantified; it is  independent of the pressure-plastic strain loading path {\color{black} and plastic deformation at pressures below  $p^d_{\vep}$}.  Thus, strain-induced PTs under compression in DAC and torsion in RDAC do not fundamentally differ. {\color{black} The yield strength of both phases is estimated using hardness and x-ray peak broadening; the yield strength in shear is not reached by the contact friction  stress  and cannot be evaluated using the pressure gradient}. Obtained results open a new opportunity for quantitative study of strain-induced PTs and reactions with applications to material synthesis and processing, mechanochemistry,  and geophysics.
 \end{abstract}

\maketitle

PTs  caused by large plastic shear deformations under high pressures play an important role in various processes in different material systems.
One of the mechanisms of a deep earthquake can be related to  the instability due to shear strain-induced PT \cite{green+burnley-1989}. Friction, wear, and surface processing (polishing, cutting, etc.) are accompanied by large plastic shear and can be optimized by involving shear-induced PTs, e.g., in Si, Ge, and strong ceramics
\cite{Patten-04}. Also,  plastic shear: (a)  drastically reduces pressure for various PTs in different material systems  \cite{Edalati-Horita-16,Blank-Estrin-2014,Levitas-Shvedov-PRB-02,Cheng-Levitas-etal-12,Perez-2009,Srinivasaraoetal-2011,Gaoetal-19}; (b) leads to formation
of new phases  which were not produced without shear \cite{Gaoetal-19,Edalati-Horita-16,Blank-Estrin-2014,levitasetal-SiC-12}; (c) substitutes   reversible PTs
        with  irreversible ones \cite{Edalati-Horita-16,Blank-Estrin-2014,Levitas-Shvedov-PRB-02}, which  allows one to use the high-pressure phases at normal pressure { in engineering applications.} These empirical results show potential in
 the development of the economic routes for  the synthesis of new and known high-pressure phases at low pressure.

The basic difference between the plastic {\it strain-induced PTs under
high pressure} and {\it pressure-induced PTs} was formulated in  \cite{levitas-mechchem-04,levitas-prb-04}. Pressure- and stress-induced PTs occur mostly by nucleation at the pre-existing defects (e.g., dislocations) below the yield. Strain-induced PTs occur by nucleation at new defects generated during plastic flow. Strain-induced PTs require completely different thermodynamic and kinetic treatment and experimental characterization.
Theoretical and computational studies of  the effect of stress tensor and plastic strain on PTs were performed
at the atomic scale \cite{Levitasetal-PRL-17,Zarkevichetal-18}, at the nano- and at the microscale using corresponding phase field approaches \cite{Javanbakht-Levitas-PRB-16,Levitas-etal-PRL-18}, at microscale by developing PT kinetics  \cite{levitas-mechchem-04,levitas-prb-04}, and at macroscale by studying behavior of  a sample under compression in DAC and compression and torsion in RDAC \cite{Levitas-Zarechnyy-PRB-RDAC-10,Feng-Levitas-JAP-16,Feng-Levitas-MSEA-16,Feng-Levitas-Mehdi-MSEA-18,Gaoetal-19}.
However, {\it  no quantitative experimental characterization of strain-induced PTs for any material and, consequently, no verification of the main hypotheses and results of simulations has been reported hitherto}.

Strain-induced PTs under high pressures are usually studied
using high-pressure torsion in RDAC \cite{Blank-Estrin-2014,Cheng-Levitas-etal-12,Gaoetal-19,levitasetal-SiC-12} or { metallic/ceramic}  Bridgman anvils
\cite{Zilbershtein-75,Edalati-Horita-16,Blank-Estrin-2014,Perez-2009,Srinivasaraoetal-2011,Edalatietal-MSEA-2009,Zhilyaevetal-MSEA-2010,Zhilyaevetal-MSEA-2011}.
As simulations show \cite{Levitas-Zarechnyy-PRB-RDAC-10,Feng-Levitas-JAP-16,Feng-Levitas-MSEA-16,Feng-Levitas-Mehdi-MSEA-18}, stress, strain, and concentration of the high-pressure phase fields are very heterogenous and vary during loading.
Though radial distribution of pressure, $p$, is measured  in DAC and
RDAC \cite{Li-etal-PNAS-17,Hemley-Science-97,Blank-Estrin-2014}, {\it fields of plastic strain and concentration of high-pressure phase, $c$, required for finding the kinetic equation, have never been reported}.

High-pressure torsion  in { metalic/ceramic} anvils does not allow for in situ studies of heterogeneous fields.
Pressure is estimated as force divided by area of the sample, which may differ from the maximum pressure in a sample by a factor of 3 and more \cite{Feng-Levitas-MSEA-16,Feng-Levitas-Mehdi-MSEA-18}. Plastic shear is  evaluated using  linear distribution for the torsion problem, which significantly differs from much more precise numerical solutions  \cite{Feng-Levitas-Mehdi-MSEA-18}. That is why PT-kinetics determined with such simplifications
(e.g., in \cite{Edalatietal-MSEA-2009}) is far from being correct.

Here, we report the first in situ quantitative XRD study of strain-induced $\alpha-\omega$  PT in ultra-pure
strongly plastically pre-deformed Zr (for which strain hardening is saturated and some critical microstructure is reached) under compression in DAC and torsion under fixed force in RDAC. {\color{black}   We hypothesize that since saturated strain hardening  drastically simplifies the plasticity theory \cite{Levitas-book-96,Levitas-etal-NPJ-CM-19}, it may also drastically simplify the kinetics of  strain-induced   PTs, and this is the best initial state to start with.  We confirm this hypothesis experimentally and obtain a number of important quantitative results on radial distribution of the in situ measured fields, the main kinetic regularities for the PT, yield strength of $\alpha$ and $\omega$ phases, and contact friction.}

 {\it Strain-controlled kinetic equation} derived based on nanoscale mechanisms  \cite{levitas-mechchem-04,levitas-prb-04}, after simplifications and modification, is presented
 for $ p_\alpha  > p_\vep^d  $  as  {\color{black}
\bey
\frac{d   c}{d   q}   =k B \frac{(1-c)}{c+(1-c) B}  \frac{p_\alpha (q) -  p_\vep^d}{p_h^d  -  p_\vep^d};         \quad   B=   \left(\frac{\sg_y^{\omega}}{\sg_y^{\alpha}}\right)^w .  
\label{l-g-4}
\eey
} Here $ q $  is the accumulated plastic strain defined by $\dot{q}= (2/3 d_p^{ij}  d_p^{ij})^{0.5}$, $d_p^{ij}$ are components of plastic deformation rate, $p_\vep^d$ and $p_h^d$ are the minimum pressures for the strain-induced and pressure-induced  $\alpha-\omega$ PT, respectively,
$k$ is a parameter, and $p_\alpha (q)$ is the pressure in the $\alpha$ phase - accumulated plastic strain loading path,  which  material particle undergoes.
Generally, Eq.(\ref{l-g-4})
  includes the term for the reverse PT 
  \cite{levitas-mechchem-04,levitas-prb-04}.  Since we obtained that  strain-induced  reverse  $\omega-\alpha$ PT in Zr does not occur, this term was dropped.
   Another simplification is in the choice of linear functions in terms of $c$ and $p_\alpha$.
Also, we changed  pressure in mixture $p$ to $p_\alpha $ in  Eq.(\ref{l-g-4})  because PT occurs in $\alpha$-Zr and we do not need to assume the same pressure in both phases as in \cite{levitas-mechchem-04,levitas-prb-04}, because we can measure them.

{\it Material.} Zr and its alloys have applications in the aerospace, nuclear, and biomedical industries due to their mechanical strength, stiffness, resistance to degradation and corrosion,
and light weight. PTs in Zr are studied in numerous papers to understand basic features of PT in solids in general and more specifically in the group IV transition metals.
 We do not know any in situ XRD studies of strain-induced PT in  Zr
 and any studies of strain-induced PT in ultra-pure Zr.
 Note that in all known studies of commercially pure Zr (99.98\% trace metals basis), Zr contained 45000 ppm  of hafnium, which is very difficult to remove. Our Zr sample, purchased from Ames Laboratory, is ultra-pure as it contains $<$55 ppm of Hf, similar to Zr in \cite{Zhao-Zhang-2007,Velisavljevic-etal-JPCM-11}.  Samples were plastically pre-deformed through cold rolling from initial thickness $h_{in}$=1.25 $mm$   down to $h_0$=140 $\mu m$ or  90 $\mu m$ (i.e.  $q\simeq ln (h_{in}/h_{0})$=2.19 or 2.63). After such large plastic straining strain hardening is saturated {\color{black} and measured Vickers hardness and consequently} the yield strength does not change, i.e., some critical microstructure is reached \cite{Levitas-book-96,Edalati-Horita-16}.
 To  study the effect of initial state, one sample was subsequently annealed at 650$^o$C for 2 hours.  Several compression and shear experiments (runs) were performed, both using steel gasket and without gasket. Experiments,  {\color{black}  measurements details, error estimates,  and experimental runs} are described in supplemental material \cite{supplemental}.
%
%

Under {\it hydrostatic loading},
 $\alpha-\omega$  PT started at pressure $p_h^d=5.4$ GPa and finished at 6.6 GPa.
 The third order Birch-Murnaghan equation of state fitting on pressure-volume data   for $\alpha$ and $\omega$ phases provides: initial volume $V_0=$23.272(2) $\AA^3$ and 22.870(8) $\AA^3$ (per formula unit); initial bulk modulus $K_0=$ 92.2 GPa and 102.4 GPa, and pressure derivative $K'=$ 3.43 and 2.93, respectively. The
 $\omega-\beta$  PT started at 34.6 GPa and finished at 35.5 GPa. Reverse   $\beta-\omega$  PT started at 34.2 GPa and finished at 32.9 GPa;  $\omega$-Zr retained at ambient pressure on complete pressure release. All   results for  hydrostatic loading are very close to those for commercially pure Zr \cite{Banerjee-2007,Perez-2009,Zhilyaevetal-MSEA-2011}.

\begin{figure}[h!]
\includegraphics[width=\linewidth]{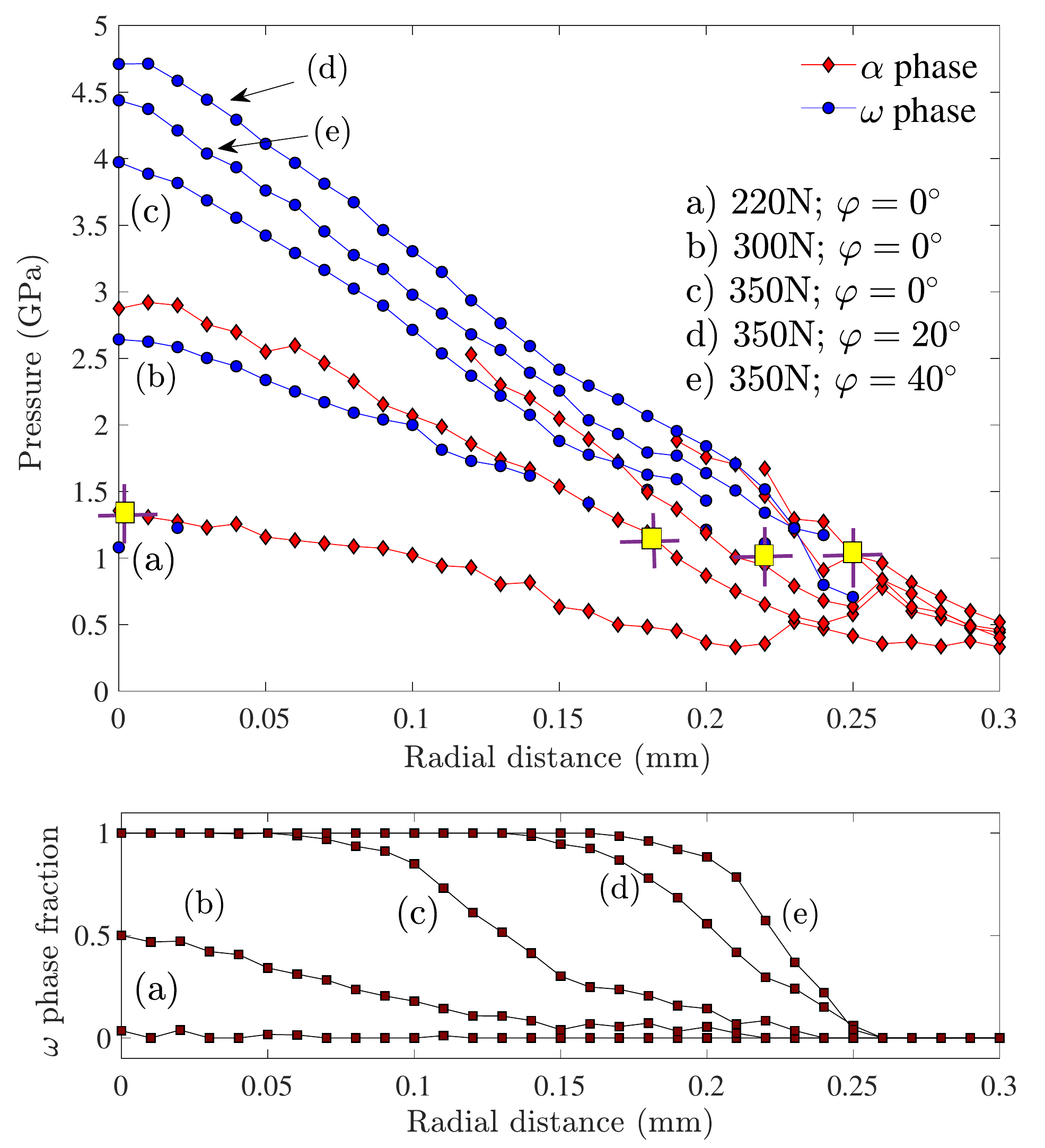}
\caption{Radial distribution of pressure in $\alpha$ and $\omega$ phases of Zr and volume fraction of $\omega$-Zr phase { in experimental run $ \sharp$1}. Inset shows axial force and rotation angle of an anvil $\varphi$. {\color{black} Yellow squares correspond to the minimum pressure $p_{\vep}^d= 1.2 \pm 0.2$ GPa  at which $\omega$-Zr was first detected at the center and periphery of the sample.}}
\label{fig:radial p-c distribution}
\end{figure}

{\color{black}  {\it The yield strength} calculated using measured saturated hardness was $\sg_y^{\alpha}$=0.70$\pm$ 0.01 GPa and $\sg_y^{\omega}$=1.37$\pm$ 0.01 GPa; using X-ray peak broadening  $\sg_y^{\alpha}$=0.70 $\pm$ 0.06 GPa and $\sg_y^{\omega}$=1.47$\pm$0.12 GPa, see details and comparison with literature in \cite{supplemental}. Both methods give surprisingly close values.
}

{\it Pressure distributions.} For  plastically pre-deformed sample without gasket, the radial distribution of pressure in  $\alpha$ and $\omega$ phases, pressure in mixture,  and volume fraction of  $\omega$ phase, averaged over the sample thickness for different compression/torsion stages are shown in Fig. \ref{fig:radial p-c distribution}. Pressure in $\omega$-Zr is slightly lower than in $\alpha$-Zr because of volume reduction. This difference reduces with increasing radius, because plastic strain  increases with radius and relaxes internal stresses between phases.
Surprisingly, pressure distribution in mixture does  not exhibit any visible signature of PT, like plateaus or change in slope \cite{Blank-Estrin-2014,Levitas-Zarechnyy-PRB-RDAC-10,Feng-Levitas-MSEA-16,Feng-Levitas-Mehdi-MSEA-18}. It is practically linear in the major part of a sample, like in simulations 
with equal yield strength of phases \cite{Levitas-Zarechnyy-PRB-RDAC-10}.
{\color{black}
However, the yield strength $\sg_y$ determined based on a simplified equilibrium equation combined with plastic friction and Tresca plasticity criterion, $dp/dr= -\sg_y/h$ , is $\sim$0.46 GPa for $\alpha$-Zr and grows from 0.46 GPa to 0.92 GPa during PTs \cite{supplemental}.  This  is contradictory because $\sg_y$ should not depend on plastic strain and the pressure change is quite small. Since these values are smaller than those obtained above by two different methods, we may conclude that the shear stress did not reach the yield strength in shear at the contact surface and that the contact friction should obey the Coulomb law  as the only currently existing alternative
\cite{Feng-Levitas-MSEA-16,Feng-Levitas-Mehdi-MSEA-18,Levitas-Zarechnyy-PRB-RDAC-10,Levitas-etal-NPJ-CM-19}.
The lack of visible signatures of the PT at the pressure distribution can be then explained by approximately equal friction coefficients of $\alpha$-Zr  and $\omega$-Zr.
 This eliminates the above contradiction in the evolution of pressure distribution. This also shows the issue with determining the yield strength based on the pressure gradient method when one cannot prove that the friction stress has reached the yield strength in shear.  }



\begin{figure}[h!]
\includegraphics[width=\linewidth]{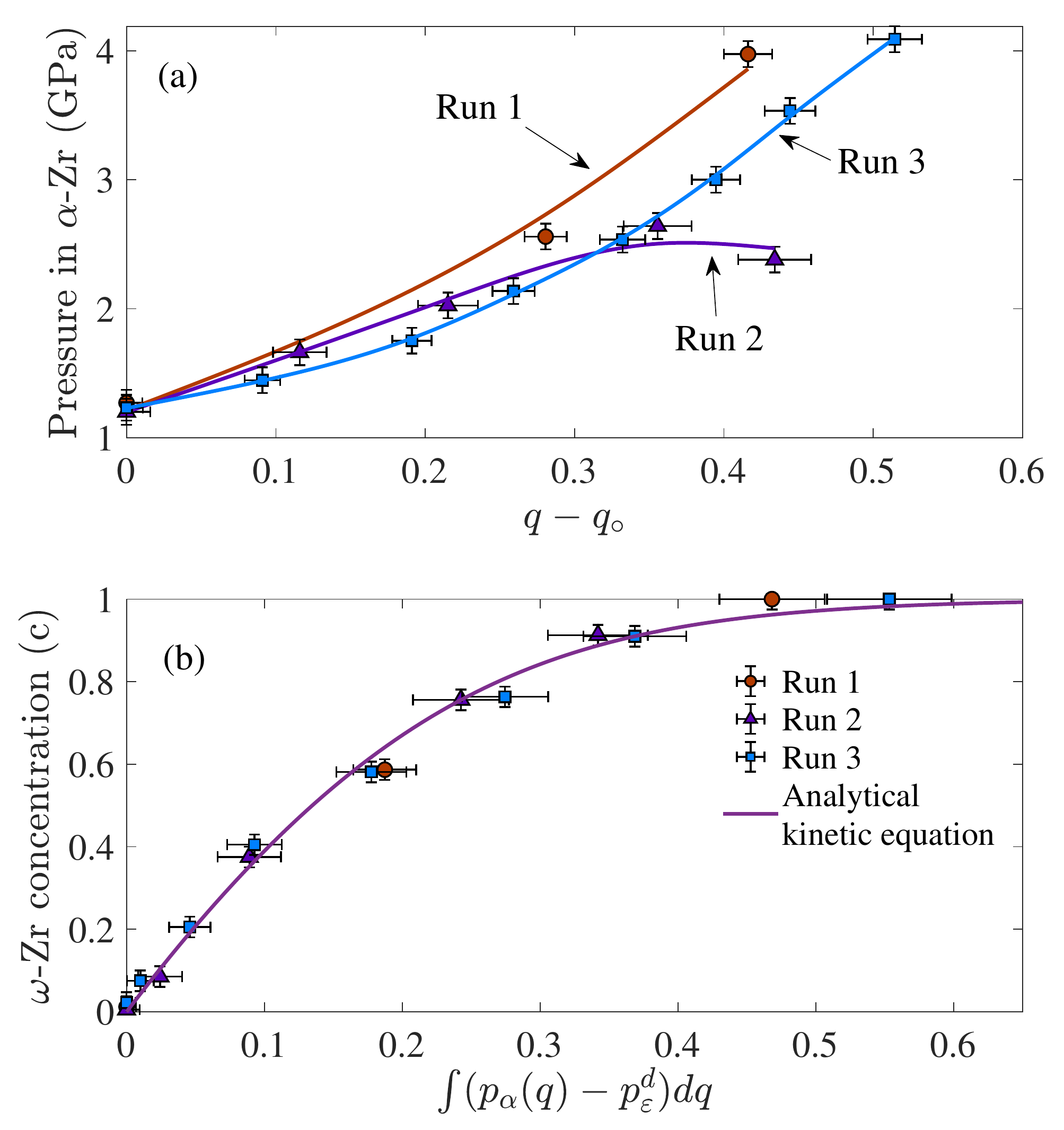}
\caption{Pressure in $\alpha$-Zr $p_\alpha$ - accumulated plastic strain $q-q_0$ loading paths for three experimental runs { described in \cite{supplemental}} (a) and kinetics of plastic-strain induced $\alpha-\omega$ PT in Zr (b).
{\color{black} Here $q_0$ corresponds to initiation of PT, i. e. to $p_\alpha= p_\vep^d$; $q_0=$0.37, 0.82, and 0.41 for runs 1, 2, and 3, respectively. Hence, the PT kinetics is independent of $q_0$. The solid line in (b) corresponds to Eq.(\ref{l-g-5})   with $k=21\pm 5$, $B=2.12$, and (for $\sigma_y^{\alpha}=0.70 $ GPa and $\sigma_y^{\omega}=1.37 $ GPa) $w=1.12$.
}}
\label{fig:kinetic eqn fitting}
\end{figure}

{\it The minimum pressure for the strain-induced $\alpha-\omega$ PT,} $p_{\vep}^d= 1.2 \pm 0.2$ GPa,  was the same
{\color{black} for all three runs and} using two methods: minimum pressure at the center of a sample during compression at which $\omega$-Zr was first detected, and based on pressure at the largest radius where $\omega$-Zr was observed, after different compression/torsion loadings  {\color{black} (Fig. \ref{fig:radial p-c distribution}a).
Thus, for strongly plastically predeformed  Zr,  plastic straining

(a) reduced the minimum PT pressure by a factor of 4.5 in comparison to hydrostatic loading, even well below than the phase equilibrium pressure of 3.4 GPa \cite{Zhangetal-2005,supplemental};

(b)  $p_{\vep}^d$ is independent of  the magnitude of plastic strain
$q_0$ below $p_{\vep}^d$ (Fig. \ref{fig:kinetic eqn fitting});}

(c) $p_{\vep}^d$ is {\it independent of compression/shear plastic stain state and its path.}
Indeed,  with torsion and without torsion,  at the center of a sample there are no shears and straining is on average unidirectional compression, but  at the periphery there are large plastic shears and multiple complex compression/shear paths.
Thus, $p_{\vep}^d$ and consequently, corresponding
physics and mechanisms are { independent of plastic stain state and its path}, which are very complex both in DAC and RDAC
\cite{Levitas-Zarechnyy-PRB-RDAC-10,Feng-Levitas-JAP-16,Feng-Levitas-MSEA-16,Feng-Levitas-Mehdi-MSEA-18}.
This also means that there is no  { advantage} in the shear mode of plastic straining, any plastic straining path produces the same effect.
Consequently, PT processes under  compression in DAC without hydrostatic medium and torsion in RDAC should be treated in the same way.
This  was postulated in \cite{levitas-mechchem-04,levitas-prb-04}, utilized in kinetic  Eq.(\ref{l-g-4}),
and used in all simulations of the processes in DAC and RDAC \cite{Levitas-Zarechnyy-PRB-RDAC-10,Feng-Levitas-JAP-16,Feng-Levitas-MSEA-16,Feng-Levitas-Mehdi-MSEA-18}, { without any experimental confirmation.}
However, in most experiments \cite{Edalati-Horita-16,Blank-Estrin-2014,Perez-2009,Srinivasaraoetal-2011,Gaoetal-19}, it was claimed that   plastic shear reduces PT pressure in comparison with uniaxial compression, while this statement was not supported by precise in situ measurements; see also our supporting results below for annealed Zr.

{\it Torsion of transformed $\omega$-Zr} was produced during reduction of load and consequently pressure.
However, even at 0.2 { GPa} at the periphery and  $\varphi$=180$^{\circ}$  reverse PT was not observed.
This is different from commercially pure Zr, for which pressure for reverse strain-induced PT was the same as for direct PT in the range of 2-2.5 GPa \cite{Zilbershtein-75,Blank-Estrin-2014}. This implies  that Hf destabilizes $\omega$-Zr at low pressure against strain-induced PT to $\alpha$-Zr.


{\it Strain-controlled kinetic equation}. Due to pressure gradient and existence of the minimum PT pressure $ p_\vep^d$, PT starts during compression at the center, and then propagates during further compression and torsion  toward periphery. Similarly, PT completes first at the center and single phase $\omega$-Zr region spreads toward edge of the culet. Note that for commercially pure Zr $\alpha-\omega$ PT does not complete even at averaged pressure of 6 GPa and 10 anvil turns \cite{Edalatietal-MSEA-2009}.

Due to impossibility to measure strongly heterogeneous fields of plastic strain in the entire sample and its large indeterminacy, we suggest to limit data for determination of the kinetic equation to the X-ray spot focused at the sample center.
At the symmetry axis, shear strains are zero.
 During compression and torsion (note that thickness $h$ reduces during torsion as well),
each point at the symmetry axis undergoes  uniaxial compression in the axial direction and expansion in the radial direction determined from the condition of plastic incompressibility, i. e., like in unidirectional compression test. Since pressure and volume fraction of $\omega$-Zr are averaged over sample thickness, plastic strain should be averaged as well.
For unidirectional compression,  $q = ln (h_0/h)$. We can assume this equation to estimate accumulated plastic strain averaged over thickness and small region within the X-ray spot.
Then determining $c$, $p_\alpha$, and $q$, we can validate Eq.(\ref{l-g-4}).
We neglect small reduction of $q$ due to elastic and transformational
strain, {\color{black} see \cite{supplemental}.}
\

All experimental points obtained in three compression and compression/torsion loadings {(runs)}
produce three different  $p_\alpha-q$ paths as shown in Fig. \ref{fig:kinetic eqn fitting}(a). They were approximated by splines and substituted in the integrated Eq.(\ref{l-g-4})   {\color{black}
\bey
c (B-1) - \ln (1-c)=   \frac{kB}{p_h^d  -  p_\vep^d} \int_{q_0}^q (p_\alpha (q) -  p_\vep^d)  dq .
\label{l-g-5}
\eey
 } Experimental points for all three paths  in coordinates $c-\int_{q_0}^q (p_\alpha (q) -  p_\vep^d)dq$ practically
 coincide (Fig. \ref{fig:kinetic eqn fitting}). They
are  close to the analytical Eq.(\ref{l-g-5})  with parameters  determined from the best fit and shown in Fig. \ref{fig:kinetic eqn fitting}.
Thus, kinetic Eq.(\ref{l-g-4})  derived based on nanoscale mechanisms in \cite{levitas-mechchem-04,levitas-prb-04} received first experimental confirmation for quite complex loading paths.

{{\it For annealed Zr}, $p_\vep^d=2.3$ GPa at the center during compression, but away from center, after much larger compression-shear plastic straining, minimum pressure at which   $\omega$-Zr is observed is the same  $p_\vep^d=1.2 $ GPa as in the center of pre-deformed Zr. Traditional interpretation of these results would be that plastic shear reduces PT pressure in comparison with plastic compression. However, since it is not true for heavily pre-deformed Zr, we suggest different interpretation. Minimum pressure $p_\vep^d$  depends on the initial microstructure
(grain size, dislocation density) and sufficient obstacles to produce strong stress concentrators, e.g.,  at dislocation pileups \cite{levitas-mechchem-04,levitas-prb-04,Javanbakht-Levitas-PRB-16,Levitas-etal-PRL-18}. After some critical preliminary plastic strain, the yield strength reaches its maximum value and does not change anymore \cite{Levitas-book-96,Edalati-Horita-16} and critical microstructure is achieved, after which $p_\vep^d$ does not change after further plastic straining. During compression of annealed Zr, pressure at the center exceeds 1.2 GPa  before plastic strain reaches critical microstructure and PT starts at 2.3 GPa, which is the   $p_\vep^d$ for the achieved plastic straining. Eq.(\ref{l-g-4}) should be generalized for pre-critical initial states by using $p_h^d (q)$ and $ p_\vep^d(q)$,
which will be done in the future work.
}

Note that transformation to $\beta$-Zr for annealed sample within steel gasket was not observed even at maximum pressure of 13 GPa and total rotation of 200$^o$. This is consistent with atomistic simulations \cite{wangetal-2011},
in which $\beta$-Zr has imaginary phonon frequencies and is dynamically unstable below 25 GPa.
This is very different from results for commercially pure Zr, where
a mixture of $\omega+\beta$  phases was reported after compression at  1 GPa   in \cite{Srinivasaraoetal-2011}, and  at 0.5 GPa  after 5 anvil turns in \cite{Zhilyaevetal-MSEA-2011}. While due to strong stress heterogeneities these numbers should be multiplied by a factor of 3 to 5 \cite{Feng-Levitas-MSEA-16,Feng-Levitas-Mehdi-MSEA-18}, still
they are much below than obtained in the current in situ experiments.

{\it In summary,}
 the first in situ quantitative characterization of plastic strain-induced $\alpha-\omega$ PT in ultra-pure Zr under high pressure in RDAC is performed utilizing synchrotron XRD.
 The evolution of the  radial distributions  of pressure in each phase and in mixture,  concentration of $\omega$-Zr, and sample thickness profile were determined. {\color{black} For strongly plastically predeformed Zr} kinetic  Eq.(\ref{l-g-4}) is confirmed and calibrated for the first time, using measurements at the sample symmetry axis,  where accumulated plastic strain $q$ can be easily evaluated.
 {\color{black} Eq.(\ref{l-g-4}) is independent of  plastic deformation at pressures below  $p^d_{\vep}$} and of the $p-q$ path.
The minimum pressure for the strain-induced $\alpha-\omega$ PT, $p_{\vep}^d=1.2$ GPa, is also independent of the
compression/shear straining path. Consequently, the strain-induced PTs under compression in DAC and shear in RDAC do not fundamentally differ in terms of kinetics, and, consequently, physical mechanisms, and modeling.
In contrast to traditional wisdom, { it is not plastic shear only but any plastic straining produces the same effect and  PT kinetics.}
One cannot claim that plastic shear in RDAC reduces PT pressure in comparison with uniaxial compression in DAC.
Plastic straining reduces PT pressure in comparison with hydrostatic loading, by a factor of 4.5 for  $\alpha-\omega$ PT in ultra-pure Zr, {\color{black} even significantly below than the phase equilibrium pressure of 3.4 GPa.}
The difference between processes in DAC and RDAC are in the $p-q$ path only. The great advantage of RDAC is that under the proper design of an experiment \cite{Feng-Levitas-JAP-16}, it allows one to run PT to completion at pressure as low as  $p_{\vep}^d$ by increasing shear. At the same time, the increase in plastic strain during compression in DAC leads to significant increase in pressure, even while it is  not necessary for PT.
{\color{black}   {  The yield strength} calculated using measured saturated hardness was $\sg_y^{\alpha}$=0.70$\pm$ 0.01 GPa and $\sg_y^{\omega}$=1.37$\pm$ 0.01 GPa; using X-ray peak broadening  $\sg_y^{\alpha}$=0.70 $\pm$ 0.06 GPa and $\sg_y^{\omega}$=1.47$\pm$0.12 GPa for $\omega$-Zr \cite{supplemental}. }
{\color{black} Surprisingly, the pressure distribution in mixture does  not exhibit any visible signature of PT, like plateaus or change in slope. It is practically linear in the major part of a sample, like in simulations 
with equal yield strength of phases \cite{Levitas-Zarechnyy-PRB-RDAC-10}.
Another contradiction was that  the yield strength $\sg_y$ determined based on pressure gradient method  grows from 0.46 GPa to 0.92 GPa during PTs, while it should not depend on plastic strain and change in pressure is relatively low. Both contradictions are eliminated by recognizing that the friction  stress did not reach the yield strength in shear at the contact surface and that the contact friction should obey the Coulomb law  as the only currently existing alternative.
The lack of visible signatures of the PT at the pressure distribution can be then explained by the approximately equal friction coefficients of $\alpha$-Zr  and $\omega$-Zr.   These results also  show risk of determination of the yield strength based on the pressure gradient when one cannot prove that the friction stress has reached the yield strength in shear.  }


In annealed Zr, $p_\vep^d=2.3$   at the center of a sample (where plastic strain is small and critical microstructure is not reached before PT) and $p_\vep^d=1.2 $ GPa at the periphery (where large plastic strain produces critical microstructure before PT), as at the center of pre-deformed Zr.
The difference between these two numbers is not due to traditional interpretation
\cite{Edalati-Horita-16,Blank-Estrin-2014,Perez-2009,Srinivasaraoetal-2011} that  plastic shear reduces PT pressure in comparison with compression, but because at the periphery much larger plastic strain produces critical microstructure before pressure reaches  $p_{\vep}^d=1.2$ GPa and at the center it does not.
Eq.(\ref{l-g-4}) should be generalized for pre-critical initial states by using $p_h^d (q)$ and $ p_\vep^d(q)$.

The obtained results allow one to transform popular qualitative discussions in literature about the effect of
plastic shear on PTs into a new quantitative field of research with applications to material synthesis and processing, mechanochemistry,  and geophysics. Current experiments can be combined with our simulations
\cite{Feng-Levitas-MSEA-16,Feng-Levitas-Mehdi-MSEA-18} in order to obtain  more advanced and precise theoretical and numerical description, extract all material parameters for more advanced models, and then find all fields
in the entire sample, even those that cannot be measured (e.g., plastic strain tensor), as it was done for deformation without PT in \cite{Levitas-etal-NPJ-CM-19}.
{ Recent advances in measurements of all components of  the stress tensor in diamond at the boundary with the sample  \cite{Hsiehetal-Science-19}  may significantly improve our boundary conditions, in particular for friction. }
The results also may lead to scientific fundamentals  for creating new, more economical deformation processes  for discovering and stabilizing high-pressure phases with novel properties, at much lower pressure.

\par\noindent\textbf{Acknowledgments:}  Supports of NSF {(MMN-1904830)}, ARO (W911NF-17-1-0225), ONR (N00014-16-1-2079), and the ISU (Vance Coffman Faculty Chair  Professorship) are gratefully acknowledged. XRD measurements were performed at HPCAT (Sector 16), APS, Argonne National Laboratory. HPCAT operations are supported by DOE-NNSA’s Office of Experimental Sciences.  The Advanced Photon Source is a U.S. Department of Energy (DOE) Office of Science User Facility operated for the DOE Office of Science by Argonne National Laboratory under Contract No. DE-AC02-06CH11357.  The authors thank Drs. D. Popov and  G. Shen (APS) for their help during measurements.

\end{document}


\beginsupplement  

\title{\textbf{Supplemental Material}\\ \vspace*{1cm}In situ quantitative  study of plastic strain-induced phase transformations under high pressure: Example for ultra-pure Zr}
\author[1]{K. K. Pandey}
\author[1,2,3]{Valery I. Levitas}
\affil[1]{\footnotesize{Department of Aerospace Engineering, Iowa State University, Ames, Iowa 50011, USA}}
\affil[2]{{\footnotesize Departments of Mechanical Engineering, Iowa State University, Ames, IA 50011, USA}}
\affil[3]{\footnotesize {Ames Laboratory, Division of Materials Science and Engineering,  Ames, IA 50011, USA}}

\date{}

\maketitle	
\begin {section}{Phase transformations in Zr}
Under quasi-hydrostatic loading Zr undergoes the   hcp ($\alpha$)$\rightarrow$ simple hexagonal ($\omega$) PT in the broad pressure range of 2-7 GPa (i.e., with large scatter from different references) and $\omega \rightarrow$ bcc ($\beta$) reversible PT at 30-35 GPa \cite{Banerjee-2007,Perez-2009,Zhilyaevetal-MSEA-2011}.
 After high-pressure torsion, a mixture of $\omega+\beta$ phases was reported at 3-6 GPa in \cite{Perez-2009}, at  1 GPa (even before torsion)  in \cite{Srinivasaraoetal-2011}, and  at 0.5 GPa  (after 5 anvil turns) in \cite{Zhilyaevetal-MSEA-2011}.
The lowest pressure for  $\alpha-\omega$ PT was  0.25 GPa  after 5 anvil turns \cite{Zhilyaevetal-MSEA-2011}.
 No other papers reported  $\omega$- and $\beta$-Zr at such low pressures. In particular, $\omega$-Zr under torsion was obtained at 2-2.5 GPa in \cite{Zilbershtein-75,Blank-Estrin-2014} and  6 GPa in \cite{Edalatietal-MSEA-2009}.
{\color{black} Since both direct and reverse $\alpha-\omega$ PT under torsion were obtained at  the same pressure in the range 2-2.5 GPa in \cite{Zilbershtein-75,Blank-Estrin-2014}, it was also claimed to be  the $\alpha-\omega$ phase equilibrium pressure. However, as it has been shown in \cite{levitas-mechchem-04,levitas-prb-04}, direct and reverse strain-induced PTs may occur at the same pressure in a broad pressure range, it is in no way related to the  phase equilibrium pressure.  Further, the  pressure was determined as a total force over total area in  \cite{Zilbershtein-75,Blank-Estrin-2014}, which may differ from the maximum pressure in a sample by a factor of 3 or more \cite{Feng-Levitas-MSEA-16,Feng-Levitas-Mehdi-MSEA-18}. Hence estimated $\alpha-\omega$ phase equilibrium pressure in  \cite{Zilbershtein-75,Blank-Estrin-2014} are not  reliable.  We do not know any in situ XRD studies of strain-induced PT in  Zr. We are also unaware of  any studies of strain-induced PT in ultrapure Zr.

 The most reliable value of the  $\alpha-\omega$  phase equilibrium pressure is  3.4 GPa \cite{Zhangetal-2005} which was obtained by extrapolating high pressure - high temperature experimental data to room temperature.
That is why we have used 3.4 GPa as the  $\alpha-\omega$  phase equilibrium pressure in the main text.}
\end{section}

\begin {section}{Sample preparation}
Ultrapure Zr samples with Hf content < 55 ppm was obtained from Ames laboratory, Ames, IA. Impurity concentrations in Zr sample are given in Table \ref{table:Zr-impurity}. The sample was cold rolled from initial thickness of 1.25 mm down to $\sim$130 to 140 $\mu m$ and 90 $\mu m$ for unconstrained and constrained compression experiments, respectively. This was to achieve saturated strain hardening through large plastic straining. For unconstrained compression experiments, 3 mm sample disks were punch cut from thinned Zr foil and were used as gasket. For constrained compression experiments,  $\sim$ 300 $\mu m$ disks were cut using laser micro-machining facility \cite{laser-cutting} at HPCAT laboratory at APS and loaded in a nearly same size  hole drilled in steel gasket pre-indented to the same thickness.

Annealed Zr samples were prepared by annealing Zr under inert {\it Ar} environment at 650$^\circ C$ for 2 hours and subsequently cooling to ambient temperature at a rate of 100$^\circ C$ per hour.
\end {section}
\begin{table}[h!]
\centering
\caption{Chemical analysis of Zr sample}
\begin{tabular}{c c| c c| c c}
 \hline
 \hline
Element & Impurity in PPM & Element & Impurity in PPM& Element & Impurity in PPM  \\
 \hline
 Al & <20 &B & <0.25&C & 32\\
 Cd & <0.25&Co & <10&Cr & <50\\
 Cu & <25&Fe & < 50&H & <16\\
Hf & <55&Mn & <25&Mo & <10 \\
 N & <20&Nb & <50&Ni &<35\\
O & 70&P &<3&Pb& <25 \\
 Si & <10&Sn & <35&Ta & <100\\
Ti & <25&U & <1&V& <25\\
W & <30& & & & \\

 \hline
\end{tabular}
\label{table:Zr-impurity}
\end{table}
\begin {section} {Rotational diamond anvil cell (RDAC)} \label{sectionRDAC}
RDAC is similar to conventional piston-cylinder diamond anvil cell (DAC), but with additional degree of freedom of relative rotation of diamond anvils with respect to each other (see Fig. \ref{fig:rdac-image}).
\
\begin{figure}[h!]
\begin{center}
\includegraphics[width=0.8\linewidth]{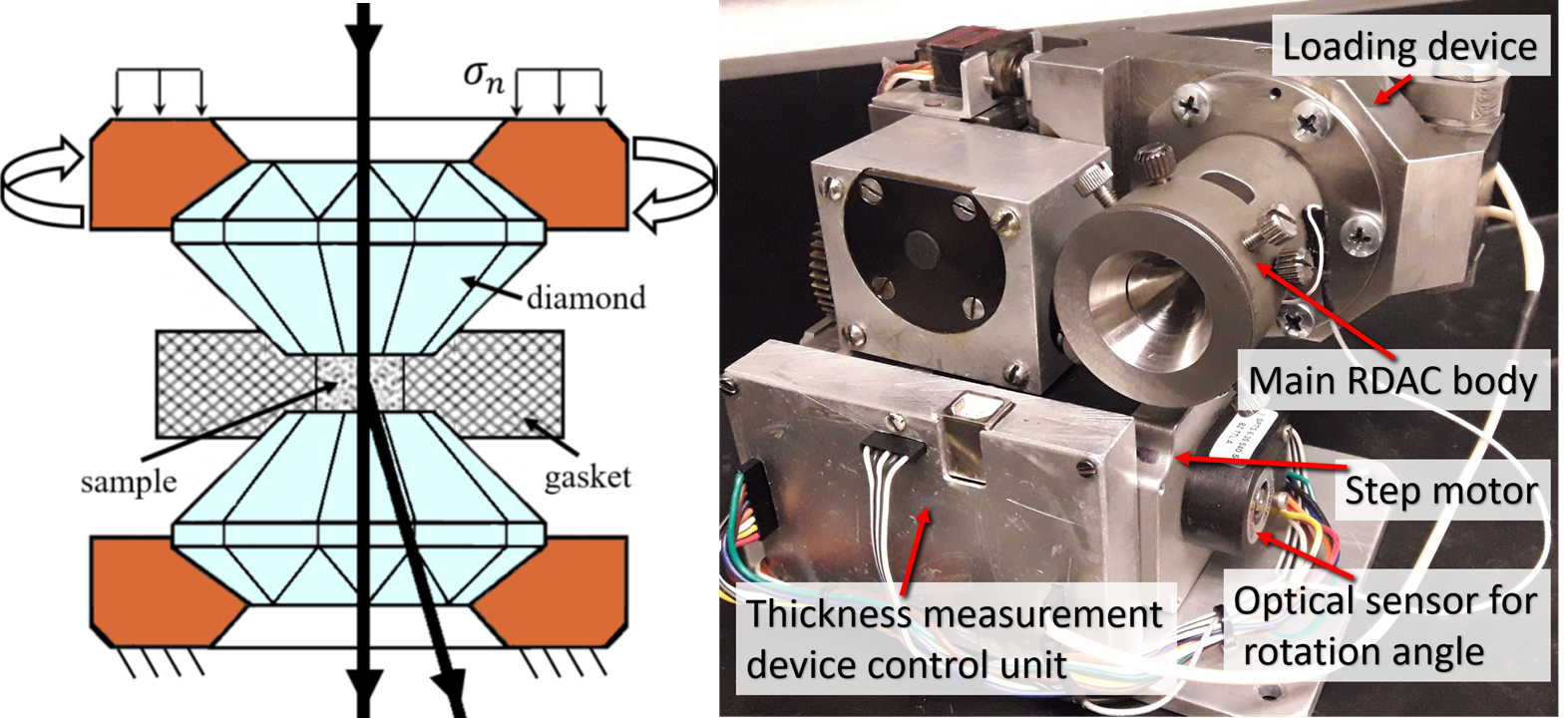}
\caption{Schematics of RDAC and its motorized mechanism. Sample is first compressed, like in traditional DAC, and a large shear is then induced by rotating one of the anvils.  }
\label{fig:rdac-image}
\end{center}
\end{figure}

Our RDAC has a motorized mechanism to mechanically apply load and/or torque at fixed load using a single stepper motor (see Fig. \ref{fig:rdac-image}). A servo-motor has been used to switch between load or shear mode by engaging the stepper motor to either compression or torsion mechanisms. The controller unit attached to the RDAC can be  connected to  a PC through USB port  and  can be remotely operated.
\end {section}

\begin {section}{Experimental details}\label{experimentaldetails}
{\color{black}Several experimental runs, prescribing different complex compression/torsion paths to plastically pre-deformed and annealed Zr samples, were performed as described in section \ref{experimentalrun}}. For unconstrained and constrained non-hydrostatic experiments, a Zr sample, with or without a steel gasket were loaded in RDAC.

For hydrostatic high pressure experiments, small Zr chips of size  ~ 20$\mu m$, obtained through diamond filing Zr sheet, were loaded in symmetric type DAC with He as a pressure transmitting medium and ruby balls as a pressure marker.

In situ XRD experiments were performed at 16-BM-D and 16-ID-B beamlines at HPCAT sector at Advanced Photon Source employing focused monochromatic x-rays of wavelength 0.3108 $\AA$ and 0.4056 $\AA$ respectively and size  $\sim$ 6$\mu m$ x 5$\mu m$ (full width at half maximum (FWHM)).  At each load - rotation angle condition, the sample was radially scanned over the entire culet diameter (500 $\mu m$) in steps of 10 $\mu m$ and 2D diffraction images were recorded at Perkin Elmer flat panel detector at 16-BM-D beamline and by the PILATUS 1MF detector at 16-ID-B beamline.

\begin{figure}[h!]
\vspace{-3mm}
\includegraphics[width=\linewidth]{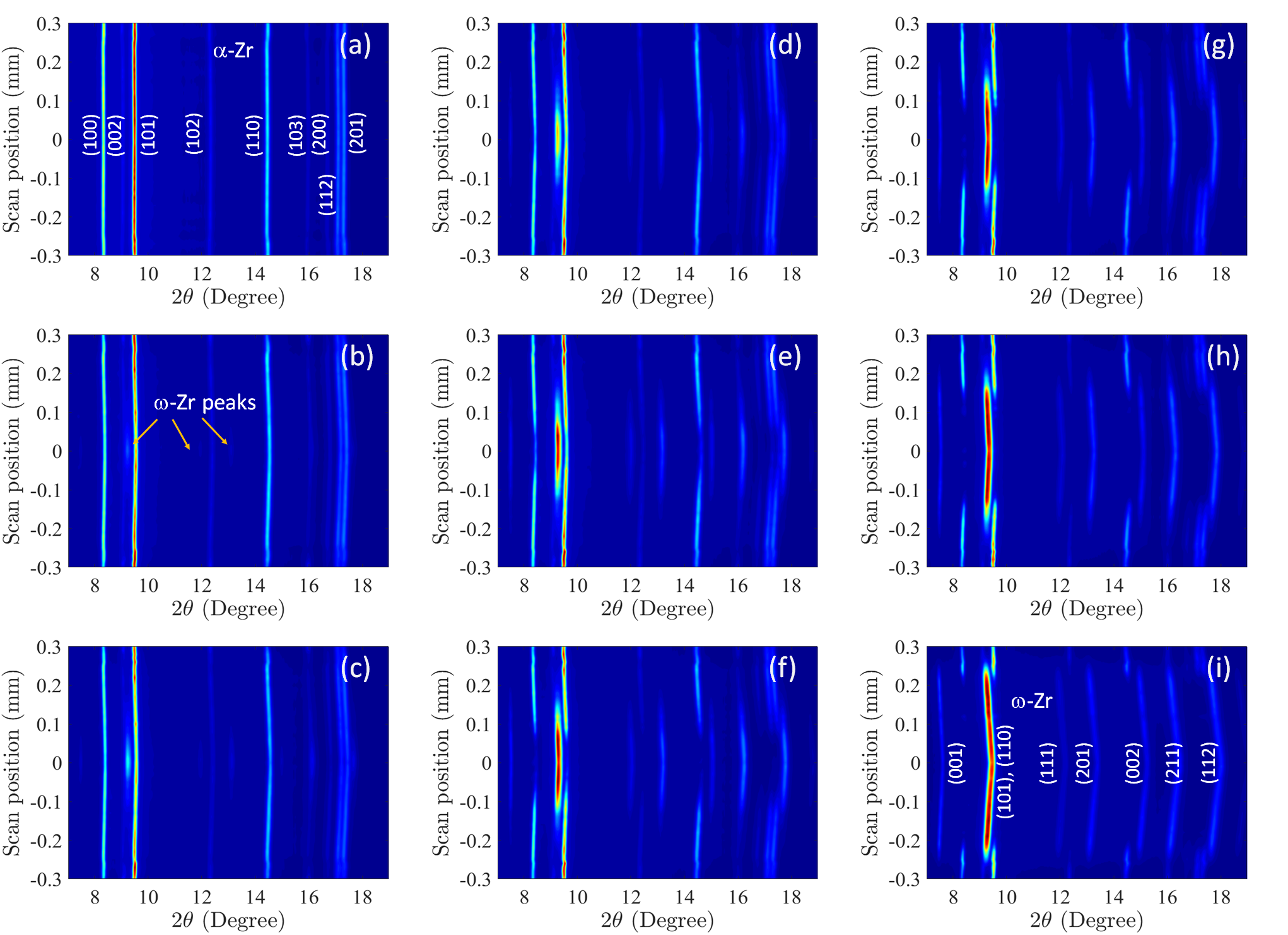}
\caption{Contour plots of XRD patterns across scanning position at various loads (a)150 $N$;(b) 200 $N$;(c) 220 $N$;(d)240 $N$;(e)260 $N$;(f)280 $N$;(g) 300 $N$;(h) 350 $N$;(i) 500 $N$;}
\label{fig:XRDcontour}
\vspace{-3mm}
\end{figure}

2D diffraction images were converted to a  1D diffraction pattern using FIT2D software \cite{fit2d1,fit2d2} and subsequently analyzed through Rietveld refinement \cite{Rietveld-1969,Young-1993} using GSAS II \cite{gsas2} and MAUD \cite{MAUD} software for obtaining lattice parameters, phase fractions and texture parameters of both $\alpha$ and $\omega$ phases of Zr. Fig. \ref{fig:XRDcontour} shows {\color{black} some representative} contour plots of XRD patterns  across scan position at various loads in one of the unconstrained compression runs. Diffraction peaks of $\omega$-Zr first emerge at the center of culet, as can be clearly seen in Fig. \ref{fig:XRDcontour}, and then spread across culet diameter. The systematic shift of $\alpha$ and $\omega$-Zr diffraction peaks across the culet is indicative of heterogeneous pressure distribution.

{\color{black} Finally, diffraction data at the symmetry axis for all load conditions were used for quantitative analysis of the kinetics of plastic strain induced $\alpha -\omega$ phase transition in Zr. For this purpose, the pressure in $\alpha$-Zr and phase fraction of $\omega$-Zr were estimated as a function of accumulated plastic strain, as described in subsequent sections.}

\end {section}

{\color{black} \begin{section}{Experimental runs} \label{experimentalrun}
Three experimental runs, designated as run $\sharp$ 1, 2, and 3, were performed on  plastically pre-deformed Zr sample in RDAC, prescribing different complex compression/torsion paths. One similar compression/torsion RDAC experiment was performed on annealed sample to study the effect of initial micro-structure. In addition to this, one constrained compression run was performed on an annealed Zr sample with a steel gasket. As mentioned above, at each load/rotation  angle state, sample was radially scanned over the entire culet diameter (500 $\mu m$) in steps of 10 $\mu m$ and XRD images were recorded at each scanning point. In runs $\sharp 1$ to $\sharp 3$, 3 mm disk of $\sim$ 140 $\mu m$ thick plastically pre-deformed Zr sample was loaded in RDAC.

In run $\sharp 1$, the sample was subjected to the axial loads of 100$N$, 150$N$, 220$N$, 300$N$, 350$N$, and subsequently to the rotation of $20^\circ$ and $40^\circ$ at the fixed load of 350$N$. After this, the load was released to 100$N$ and subsequently 50$N$ and rotation of $180^\circ$ was performed before complete release to check the reverse $\omega \to \alpha$ transition.

In run $\sharp 2$,  the sample was subjected to the axial loads of 160$N$, 190$N$, 220$N$, and subsequently to the rotation of $10^\circ$, $20^\circ$, $30^\circ$, and $40^\circ$ at the fixed load of 220$N$.

In run $\sharp 3$,   sample was loaded along with gold powder, with particle size 1 to 5 $\mu m$, sprinkled over sample. The sample was subjected to axial loads of 100$N$, 150$N$, 180$N$, 190$N$, 210$N$, 230$N$, 250$N$, 270$N$, 300$N$, 320$N$, 350$N$, 370$N$, 400$N$, 430$N$, 450$N$, 500$N$, 550$N$ and 600$N$.  Presence of relatively soft gold particles reduced friction at sample-anvil contact surface in this run. This led to different $p-q$ loading paths in runs $\sharp 3$  and $\sharp 1$,  despite prescribing nearly similar axial loads, as can be seen in Fig. 2 of the main text.

For compression/torsion run on the annealed sample, a $\sim 90 \mu m$ thick 3 mm disk of annealed Zr sample was loaded in RDAC. The sample was subjected to loads of 150$N$, 180$N$, 220$N$, 250$N$, 280$N$, and then rotation up to $80^\circ$ in steps of $10^\circ$. Subsequently, load was released in steps of 200$N$, 150$N$, 100$N$, 50$N$, and then completely released.

In the constrained compression/torsion run, a $\sim$295 $\mu m$ diameter disk of  annealed Zr sample with thickness $\sim 90 \mu m$ was loaded in $\sim$ 300 $\mu m$ hole,  drilled in  steel gasket pre-indented from 250 $\mu m$ to $\sim 90 \mu m$. The sample was subjected to loads of 100$N$, 200$N$, 250$N$, 300$N$, 350$N$, 410$N$, 500$N$, 550$N$, 610$N$, 700$N$, and subsequently rotation of 10$^\circ$, 20$^\circ$, 40$^\circ$, 100$^\circ$. After this treatment, the sample was completely transformed to $\omega$ phase. The sample was subsequently subjected to further higher loads of 800$N$, 960$N$, 1060$N$, 1150$N$, and 1250$N$.

\begin{section}{Phase fraction estimation}
{\color{black} The phase fraction of $\alpha$ and $\omega$-Zr have been estimated from Rietveld refinement \cite{Rietveld-1969} of diffraction patterns.}
\begin{figure}[h!]
\vspace{-3mm}
\begin{centering}
\includegraphics[width=0.8\linewidth]{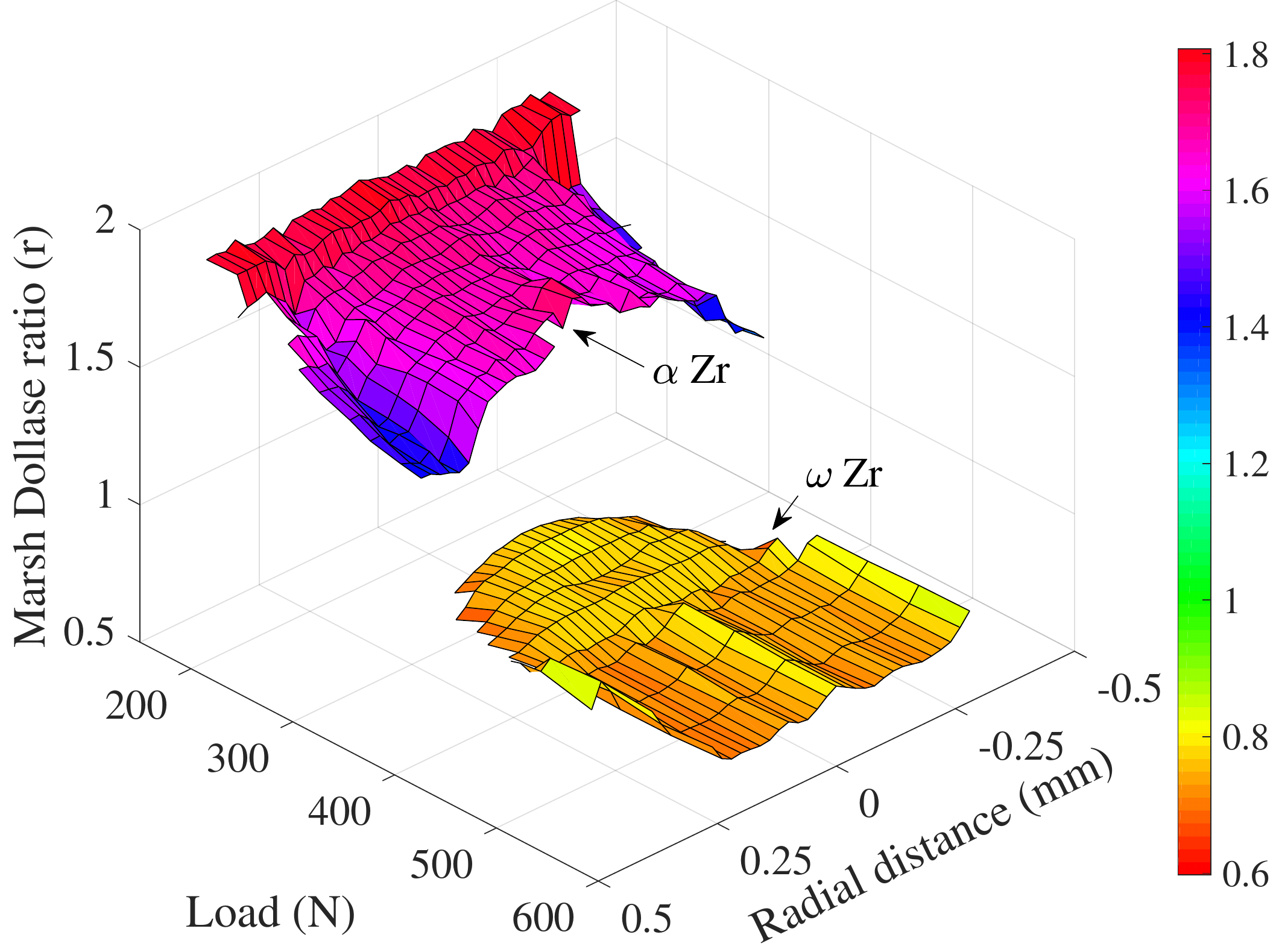}
\caption{Surface plot of Marsh Dollase ratio $r$ for (001) plane of both $\alpha$ and $\omega$-Zr across the culet diameter and at various loads}

\label{fig:MDratio}
\end{centering}
\vspace{-3mm}
\end{figure}
As the sample was highly textured due to substantial preliminary plastic deformation, the Rietveld refinement was performed with refinement of  Marsh Dollase ratio ($r$) \cite{Marsh-1932, Dollase-1986}, for (001) plane for each phase, in order to obtain correct phase fraction. Fig. \ref{fig:MDratio} shows the refined $r$ values for both $\alpha$ and $\omega$-Zr.
Initially $\alpha$-Zr is highly textured along $c$ axis parallel to the load axis and uniform across the culet diameter. At higher loads the $r$ slightly reduces  at the edges of culet. This is due to rotation of Zr grains while they move around the anvil  edges due to plastic deformation of Zr. For $\omega$-Zr, the $r$ parameter implies texturing of $c$ axis away from load direction, which is consistent with earlier reported orientation relationship between $\alpha$ and $\omega$-Zr phases after transition under high pressure\cite{Wenk-prl-2013}.

{\color{black} From Rietveld refinement of diffraction patterns, the error of phase fraction estimation  in our measurements has been estimated to be $\pm 0.05$. }

\end{section}

\end{section}
{\color{black}
\begin {section}{Pressure Estimation}
As mentioned in the main text, we need to estimate the heterogeneous field of pressure in the initial phase of a sample  for the quantitative analysis of the kinetics of plastic strain induced phase transitions. For this purpose, pressure must be estimated using elastic strains obtained from XRD measurements in the sample. Any other method based on secondary pressure markers, viz. ruby fluorescence, standard metals (e. g. gold, copper etc.), or shift of diamond Raman peak,
include additional errors. For example, (a) the pressure distribution obtained from using ruby fluorescence exhibits significant scatter due to the uncontrolled particle orientation dependence of the fluorescence; (b) also, pressure in the ruby particle and in the material under study may be different; (c) pressure is measured at the diamond-sample contact surface and does not include
averaging over the sample thickness (as it is done for the volume fraction of the high pressure phase and plastic strain); ruby particles disturb friction condition and stress distribution.

Ideally, all elastic strain components  can be estimated through measurements of  lattice strains, spatially resolved across the entire sample. However, geometrical constrains in RDAC, or any other type of available DAC, do not allow such detailed measurements. Under complex compression and torsion conditions, stress states are heterogeneous, not only across anvil diameter but across sample thickness as well. Hence, all diffraction geometries, generally used by high pressure community under non-hydrostatic stress state  \cite{AKSingh-JAP1994, AKSingh-JAP1998} suffer from some of limitations. In axial (parallel) geometry, where incident x-rays are along the load axis, XRD provides lattice parameters averaged over thickness; whereas in radial (perpendicular) geometry, where the incident x-rays are perpendicular to the load axis, obtained lattice parameters are averaged over the sample diameter. As there exists a large pressure gradient across the anvil diameter, radial geometry is ruled out for our studies as it would introduce large uncertainties in the estimation of transformation pressures and does not allow the measurement of radial distribution of pressure (and volume fraction of phases).  The only choice left is the axial geometry.  This is  a reasonable choice since heterogeneity in the normal stress components across the sample thickness are limited by the yield strength of the sample material \cite{Feng-2016,Feng-Levitas-MSEA-16,Feng-Levitas-Mehdi-MSEA-18}. The heterogeneity in pressure across the sample diameter can be estimated by scanning the sample across diameter with a focused x-ray beam and recording diffraction image at each scan position, as mentioned in section \ref{experimentaldetails}. Pressure is defined as
\bey
p=-(\sigma_{11}+\sigma_{22}+\sigma_{33})/3 ,
\label{eqn-1}
\eey
where $\sigma_{33}$, $\sigma_{11}$ and $\sigma_{22}$ are  the normal stress components along the load (vertical), radial and azimuthal directions respectively. As measurements were done in axial geometry, the angle between the diffraction plane normal and the load axis, $\psi=90^\circ-\theta_{hkl}$ ($\theta_{hkl}$ is the diffraction angle for planes with Miller indices $hkl$) for all orientations of $hkl$ planes satisfying diffraction condition, and the diffraction ring corresponding to $hkl$ plane is nearly circular. That is why  none of the procedures prescribed for radial  geometry \cite{AKSingh-JAP1994, AKSingh-JAP1998, Duffy-PRB1999, Merkel-MSMSE-2012}, where  $d_{hkl}$ are measured as a function of $\psi$, particularly at $0^\circ$, $90^\circ$ and $54.7^\circ$, can be used in this study. One possible solution for this analysis could have been using the moment pole stress model as implemented in MAUD software \cite{MAUD} and using single crystal elastic constants for Zr phases as an initial guess for refinement. However, as this implemented model is based on  elasticity theory and neglects the effect of plasticity on lattice strains, it yields inconsistent stress components and elastic constants \cite{Merkel-PRB2009}. Since plastic strains do not contribute to the Hooke's law expressed in terms of elastic strains, and during plastic deformation under normal stresses the  differential stress $t=\sigma_{33}-\sigma_{11}=\sigma_y$, we tried to overcome this issue in MAUD analysis by using the pressure dependent elastic constant of Zr phases and constraining differential stress to yield strength of Zr. However, refinement was giving arbitrary values of stress components and did not converge.

Finally, we used the following two methods to estimate pressure distribution in each phase across the anvil diameter:
\begin{enumerate}
\item{Traditional method \cite{Hemley-Science-97,Li-etal-PNAS-17} based on implementing hydrostatic equation of state of Zr phases.}
\item{Utilizing the pressure-dependent Hooke's law and elastic constants of Zr phases.}
\end{enumerate}

Both these methods give very close pressure estimates. The maximum error in pressure estimates from these methods is limited by the  yield strength of material; since Zr is a relatively soft material, these methods are reasonable within our experimental limitations. A more precise and spatially resolved estimation of all stress components and hence precise pressure distribution can be obtained through combined  finite element simulations and experiments \cite{Levitas-etal-NPJ-CM-19}, which is underway and will be reported elsewhere.

\begin{subsection}{Using hydrostatic equation of state  (Method 1)}\label{method1}

It is known that the pressure-volume relationship under hydrostatic and nonhydrostatic conditions are different
\cite{AKSingh-JAP1994, AKSingh-JAP1998, Duffy-PRB1999, Merkel-MSMSE-2012}.
However, this difference is pronounced under very high pressure, for strong materials, and when the yield strength significantly increases with pressure. Still, in \cite{Hemley-Science-97,Li-etal-PNAS-17} even at 300 GPa and 400 GPa the hydrostatic equation of state is used for pressure evaluation under strongly nonhydrostatic conditions. The main reason is that it is nontrivial to find a better way for pressure evaluation under such conditions when even shear elastic
strains are large, elasticity is strongly nonlinear and elastic moduli are unknown. Since quantitative results in the current paper are obtained for relatively low pressure, below 4 GPa, inaccuracy due to using the hydrostatic equation of state is expected to be quite low. In addition, due to small shear strains, applicability of the Hooke's law and known pressure-dependent elastic moduli for $\alpha$-Zr, these results will be verified in \ref{method2}.

We refined the diffraction pattern recorded at each scanning position using Rietveld method \cite{Rietveld-1969,Young-1993} and obtained volumetric strain. Subsequently, we used  the third-order Birch-Murnaghan equation of state of $\alpha$ and $\omega$ phases of Zr obtained from our hydrostatic experiment (see Table \ref{table:Zreos}) and estimated pressure at each scanning position in each phase. Total pressure at each scanning position was estimated based on mixture theory.
\begin{table}[h!]
\centering
\caption{Parameters in the third-order Birch-Murnaghan equation of state of Zr phases obtained from hydrostatic experiments}
\begin{tabular}{c c c c}
 \hline
 \hline
Zr phase & $V_\circ$ (per formula unit) &  $K_\circ$ &  $K'_\circ$ \\
\hline
$\alpha$-Zr & 23.272(2) $\AA^3$ & 92.2 GPa & 3.43\\
$\omega$-Zr & 22.870(8) $\AA^3$ & 102.4 GPa & 2.93\\

 \hline

\end{tabular}
\label{table:Zreos}
\end{table}
\end{subsection}
\begin{subsection}{Using pressure dependent elastic constants (Method 2)}
\label{method2}
Here we considered two main stress states under plastic deformation.

\begin{enumerate}
\item{Along the symmetry axis and at the  symmetry plane, shear stresses are zero,  and the stress component along load axis (Fig. \ref{fig:loadschematic}),
\bey
\sigma_{33}=\sigma_{11}-\sigma_y.
\label{eq-ss-at-symmetryplane}
\eey
Note that compressive normal stresses are negative.}
\item{At sample-diamond contact  surface, except in very small region exactly on the symmetry axis, friction shear stress reaches (or approaches) the yield strength in shear. For the von Mises yield condition this  results in $\sigma_{11}=\sigma_{22}=\sigma_{33}$, while for the Tresca yield condition we obtain (Fig. \ref{fig:loadschematic}) \cite{Feng-2016,Feng-Levitas-MSEA-16,Feng-Levitas-Mehdi-MSEA-18}.
\bey
\sigma_{11}=\sigma_{33}.
\label{eq-ss-at-contactplane}
\eey
Since (a) the region with reduced shear stresses near the symmetry axis is very narrow \cite{Feng-2016,Feng-Levitas-MSEA-16,Feng-Levitas-Mehdi-MSEA-18}, (b)
    there is finite size of probing x-rays ($\sim$ 6$\mu m$ x 5$\mu m$ FWHM, i. e. $\sim 10 \mu m$  for $95 \%$ beam in our experiments), and (c) the scanned position near cullet center might be slightly shifted from the symmetry axis,  such a stress state is approximately applicable at all scanning positions at the contact surface in our measurements.}
\end{enumerate}

\begin{figure}[h!]
\vspace{-3mm}
\begin{centering}
\includegraphics[width=0.8\linewidth]{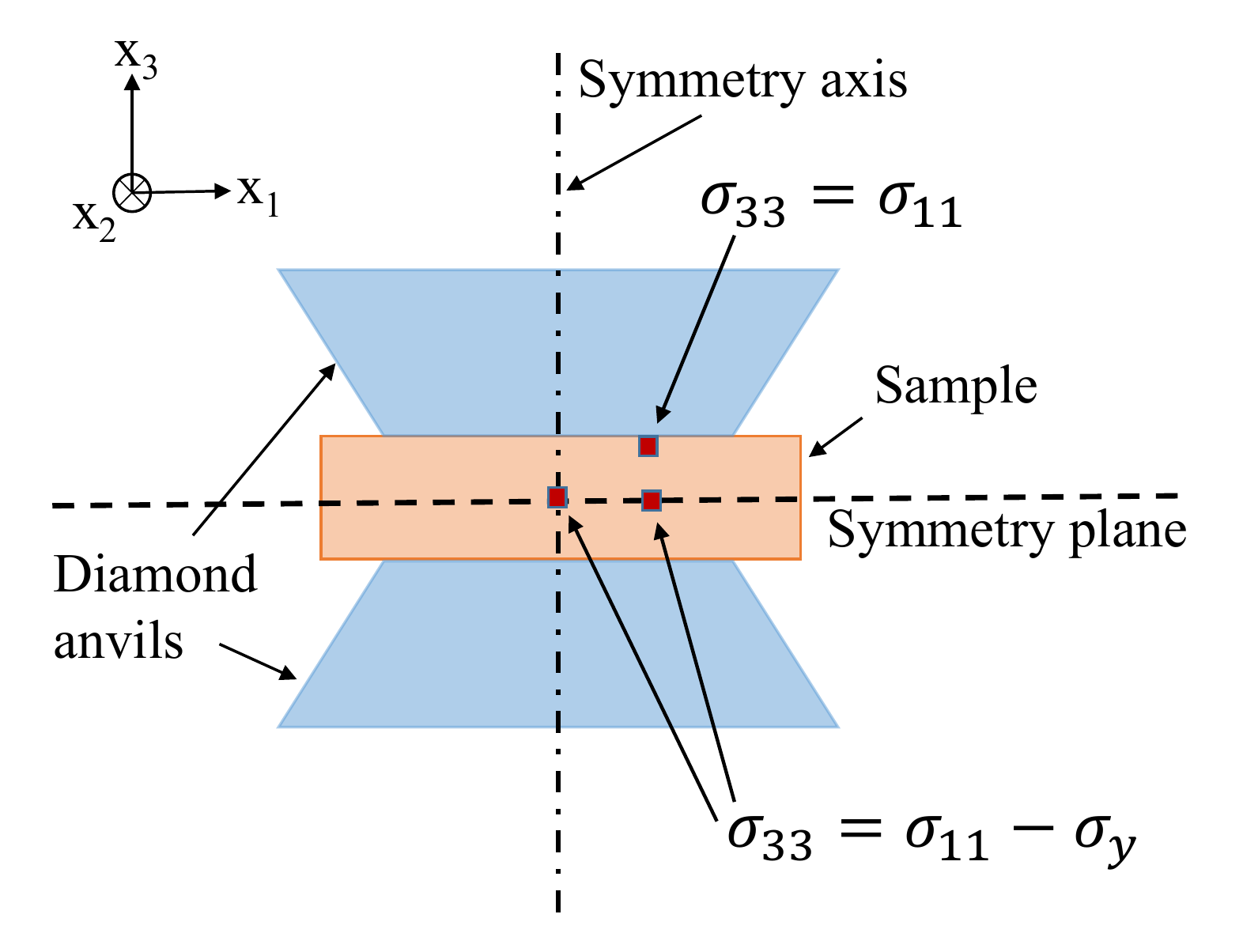}
\caption{Schematic illustration of stress states at the symmetry plane and contact surface.
}
\label{fig:loadschematic}
\end{centering}
\vspace{-3mm}
\end{figure}

Note that during torsion, the azimuthal component of the shear stress appears, which reduces differential stress   $\sigma_{33}-\sigma_{11}$ even at symmetry plane, making deviation between the normal stresses and pressure smaller.
Since this case is between  the above mentioned two extreme stress states, we have not considered it.

\begin{figure}[h!]
\vspace{-3mm}
\begin{centering}
\includegraphics[width=\linewidth]{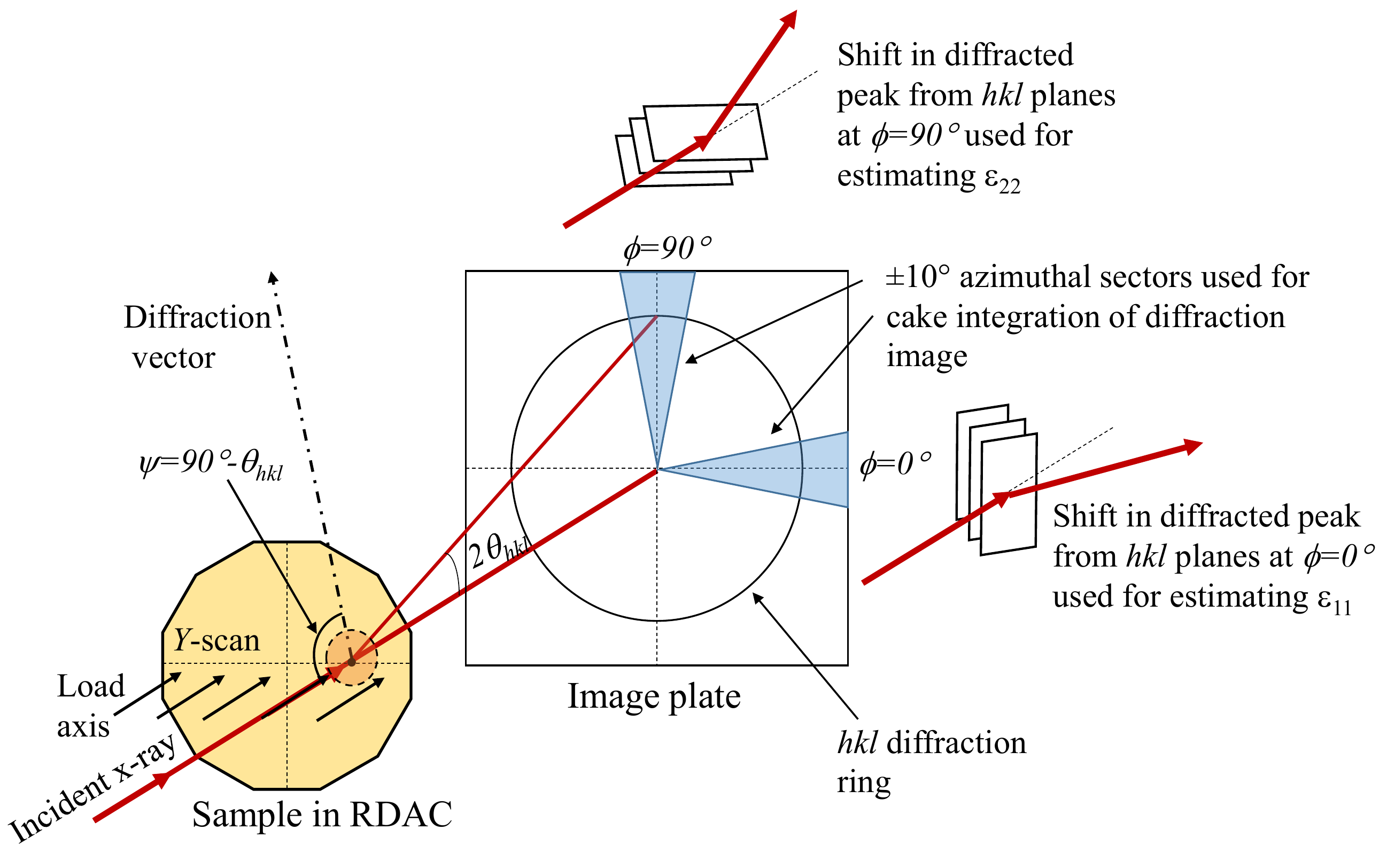}
\caption{Schematic illustration of estimation of elastic strains in radial ($\varepsilon_{11}$) and azimuthal ($\varepsilon_{22}$) directions at each scanning position.
}
\label{fig:method2schematic}
\end{centering}
\vspace{-3mm}
\end{figure}

 In axial geometry the diffraction condition is satisfied mostly for those planes which are nearly parallel (plane normal perpendicular) to the load axis. Hence the observed shifts in diffraction peaks can be practically used to estimate strains in radial and azimuthal directions viz. $\varepsilon_{11}$ and $\varepsilon_{22}$ (Fig. \ref{fig:method2schematic}). Ideally, the angle between the load axis and diffraction vector ($\psi$) should be equal to $90^\circ$ to estimates these strain components. However, since this is not possible in axial geometry, we can use the diffraction peak with smallest diffraction angle, $\theta$. In our experiments for $\alpha$-Zr, (100) diffraction peak appears at $\theta=3.18^\circ$ for used x-rays ($\lambda=3.1088 \AA$) at ambient pressure. This corresponds to $\psi=86.82^\circ$ and may be practically used for  estimation of strain components $\varepsilon_{11}$ and $\varepsilon_{22}$.  Note that (100) peak corresponds to $a$ lattice parameter and $c$-axis of $\alpha$-Zr is predominately aligned along the loading direction as per our texture analysis. Then using the measured strain components $\varepsilon_{11}$ and $\varepsilon_{22}$ and the stress conditions \ref{eq-ss-at-symmetryplane} and \ref{eq-ss-at-contactplane}, all principal stress components and hence pressure can be iteratively estimated employing linear elasticity with pressure-dependent single crystal elastic constants $C_{ij}$  of $\alpha$-Zr (Table \ref{table:Zrelasticconstants}).

\begin{table}[h!]
\centering
\caption{Elastic constants and their pressure derivatives for $\alpha$-Zr \cite{Fisher-1970}.}
\begin{tabular}{c c c c c c c}
 \hline
 \hline
$\alpha$-Zr& $C_{11}$ &  $C_{33}$ &  $C_{12}$ &  $C_{13}$ &  $C_{44}$ &  $C_{66}$ \\
\hline
$C_{ij}$ (GPa) & 143.68 &  165.17 &  73.04 &  65.88 &  32.14 &  35.32 \\
$dC_{ij}/dP$& 3.93 &  5.49 &  3.42 &  4.25 &  -0.22 &  0.26 \\

 \hline

\end{tabular}
\label{table:Zrelasticconstants}
\end{table}

The Hooke's law for hexagonal crystal for normal stresses and strains can be presented in the form
\bey
\begin{bmatrix} \sigma_{11} \\ \sigma_{22} \\ \sigma_{33} \end{bmatrix}=\begin{bmatrix} C_{11} & C_{12} & C_{13} \\ C_{12} & C_{11} & C_{13} \\ C_{13} & C_{13} & C_{33} \end{bmatrix}
\begin{bmatrix} \varepsilon_{11} \\ \varepsilon_{22} \\ \varepsilon_{33} \end{bmatrix}.
\label{eq-ss-a}
\eey

{\it At the symmetry plane}, assuming known (measured) $\varepsilon_{11}$ and $\varepsilon_{22}$ and the yield condition
$\sigma_{33}=\sigma_{11}-\sigma_y$, we can resolve Eqs. (\ref{eq-ss-a}) for
strain component $\varepsilon_{33}$ as
\bey
\varepsilon_{33}=\frac{(C_{11}-C_{13})\varepsilon_{11}+(C_{12}-C_{13})\varepsilon_{22}-\sigma_y}{(C_{33}-C_{13})}.
\label{eq-e33s-a}
\eey
Substituting all strain components in Eqs. (\ref{eq-ss-a}), we can determine all stress components. The  pressure can be evaluated using  Eq. (\ref{eqn-1}) at all scan positions.

{At sample-anvil contact surface}, using      $\sigma_{11}=\sigma_{33}$,
we can again resolve Eqs. (\ref{eq-ss-a}) for
strain component $\varepsilon_{33}$ as
\bey
\varepsilon_{33}=\frac{(C_{11}-C_{13})\varepsilon_{11}+(C_{12}-C_{13})\varepsilon_{22}}{(C_{33}-C_{13})}.
\label{eq-e33c-a}
\eey
Then again all stress components and pressure can be found using Eqs. (\ref{eq-ss-a}) and  (\ref{eqn-1}).

Note that the above procedure was performed iteratively. At each iteration step, elastic constants were updated using their pressure dependence and based on pressure estimated in the previous iteration, until convergence is reached.

Finally, pressures estimated at symmetry plane and contact surface with above described stress states were averaged to find mean pressure across sample thickness at each scanning position.

Similar procedure can be repeated for $\omega$-Zr. However, since experimental values of pressure dependent elastic constants for $\omega$-Zr are not available and elastic constants determined by atomistic simulations at 0 K \cite{Hao-2008} might give inaccurate  estimates at ambient temperature,
pressures were not estimated for $\omega$-Zr using this method.

Obtained pressure distributions for $\alpha$-Zr  using methods 1 and 2, are compared in Fig. \ref{fig:pdcomparision} (a) and (b), respectively at a few representative load/shear conditions in run $\sharp 1$ and run $\sharp 2$.
Difference, which characterizes error in pressure determinations does not exceed 0.2 GPa.
Besides, there are errors involved in the hydrostatic equation of state (bulk modulus and its pressure derivative) and the yield strengths of Zr phases. So, errors in  pressure has been considered as the convolution of average difference in the pressures estimated using methods 1 and  2, error in equation of state and error in yield strengths of Zr phases. Estimated error in pressure of $\alpha$-Zr, $\Delta p$, is   $\sim \pm 0.2$ GPa.

\begin{figure}[!htb]
    \centering
    \begin{minipage}{.5\textwidth}
        \centering
        \includegraphics[width=\linewidth]{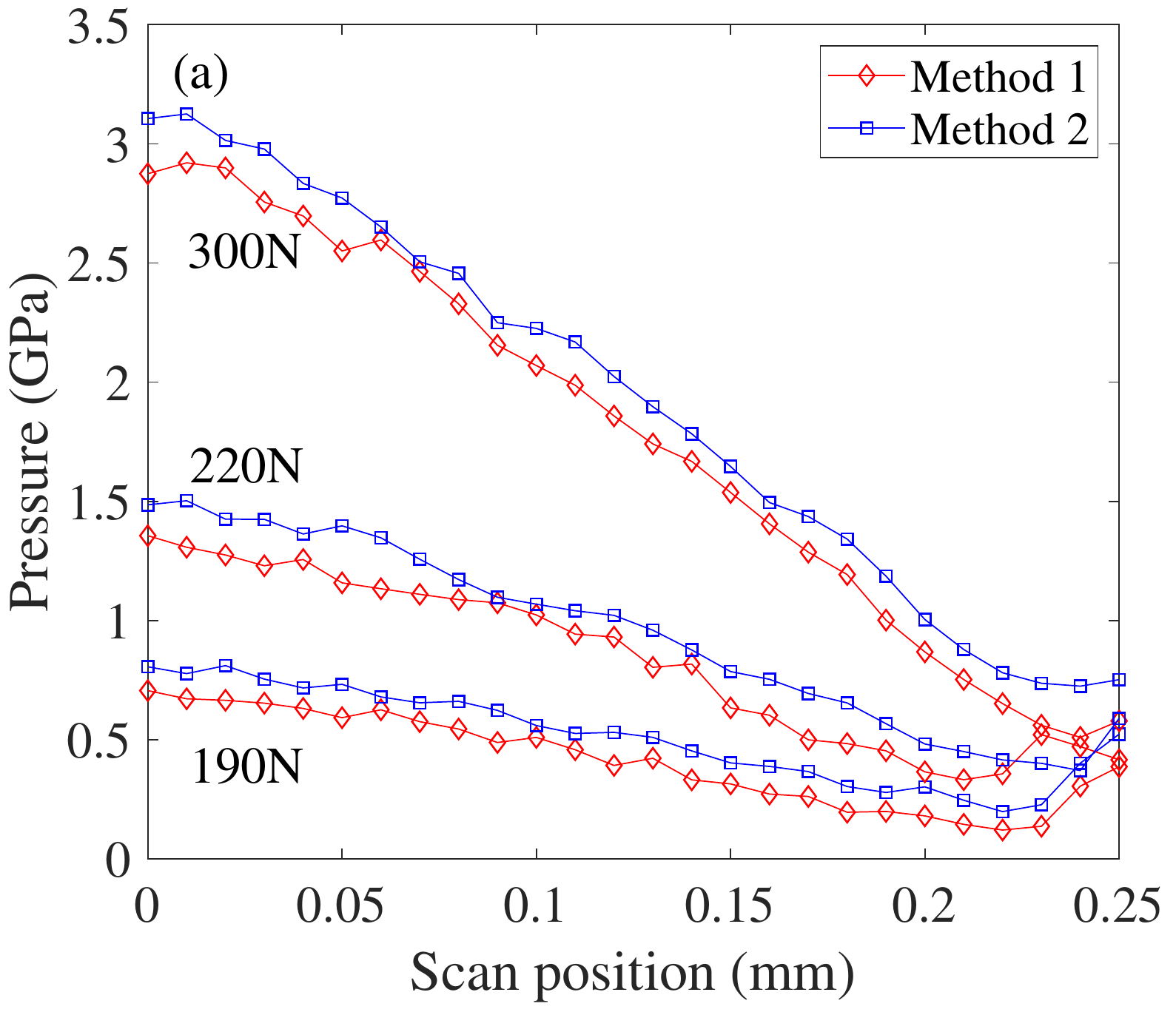}

    \end{minipage}%
    \begin{minipage}{0.5\textwidth}
        \centering
        \includegraphics[width=\linewidth]{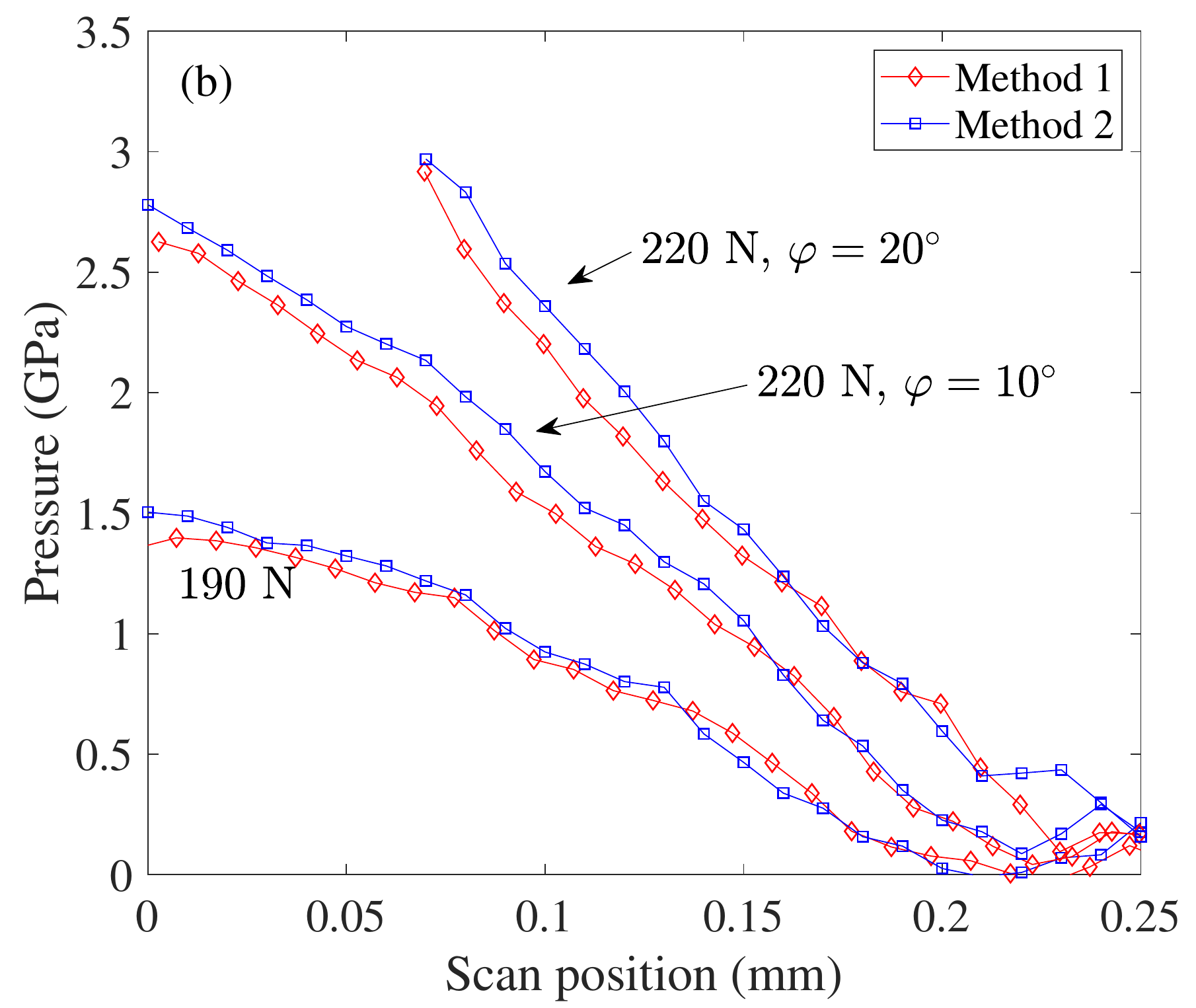}

    \end{minipage}
    \caption{Comparison of pressure distribution in $\alpha$-Zr obtained using method 1  and method 2 for (a)  run $\sharp 1$ and (b) run $\sharp 2$. In run $\sharp 2$, at 220 N load and $20^\circ$ torsion, Zr significantly transformed to $\omega$ phase up to the radial distance of 0.07 mm from culet center.  }
    \label{fig:pdcomparision}
\end{figure}

\end{subsection}

Since difference between two methods is small and within the experimental precision of pressure estimation, we use  method 1 (i.e., like in  \cite{Hemley-Science-97,Li-etal-PNAS-17}) for  quantification of kinetics of $\alpha \rightarrow \omega$ phase transition in Zr.

\end{section}

\begin{section} {Yield strength estimation} \label{yieldstrength}

\begin{subsection} {Utilizing hardness measurement} \label{hardness}

Yield strength of $\alpha$-Zr  was estimated from Vickers hardness \cite{Vickers} measurements on the initial plastically pre-deformed sample using the LECO LM 247AT micro-indentation hardness tester  at Metallography Laboratory at Iowa State University.
The tests were performed with an indentation load of 500g and dwell time of 13 sec, which was above the load and dwell time for which hardness becomes independent of these parameters. Measured maximum (saturated)   hardness after rolling from thickness 1.25 mm down to  90 $\mu m$ was 196$HV$ (1.92  GPa).
The yield strength $\sg_y = k \, HV$, with $k$ in the range from 1/3 to 0.386.  We choose the mean value $k=0.36$ and obtain $\sigma_y =0.7$ GPa.

These values of maximum hardness and the yield strength for  $\alpha$-Zr are in agreement with the corresponding values of hardness in the range of  2 - 2.4 GPa and the yield strength in the range of 0.70 - 0.74 GPa  \cite{Zhilyaevetal-MSEA-2010,Yang,Cao} for Zr of different purities.

The $\omega$-Zr   saturated value of $HV$ after high-pressure torsion at 6 HPa was  around 3.8 GPa \cite{Edalatietal-MSEA-2009}, which correspond to the yield strength of 1.37 GPa.

For annealed $\alpha$-Zr, the measured hardness was  95$HV$ (0.93 GPa), which is   close to   1 GPa in \cite{Zhilyaevetal-MSEA-2010}.
Such a hardness corresponds to $\sigma_y =0.34 $ GPa, but the error in connecting hardness and the yield strength here is larger, because of strain hardening. Direct measurements of $\sigma_y$ in tensile test gives $\sigma_y =0.32 $ GPa  \cite{Yang}.
\end{subsection}

\begin{figure}[h!]
\begin{centering}
\includegraphics[width=0.6\linewidth]{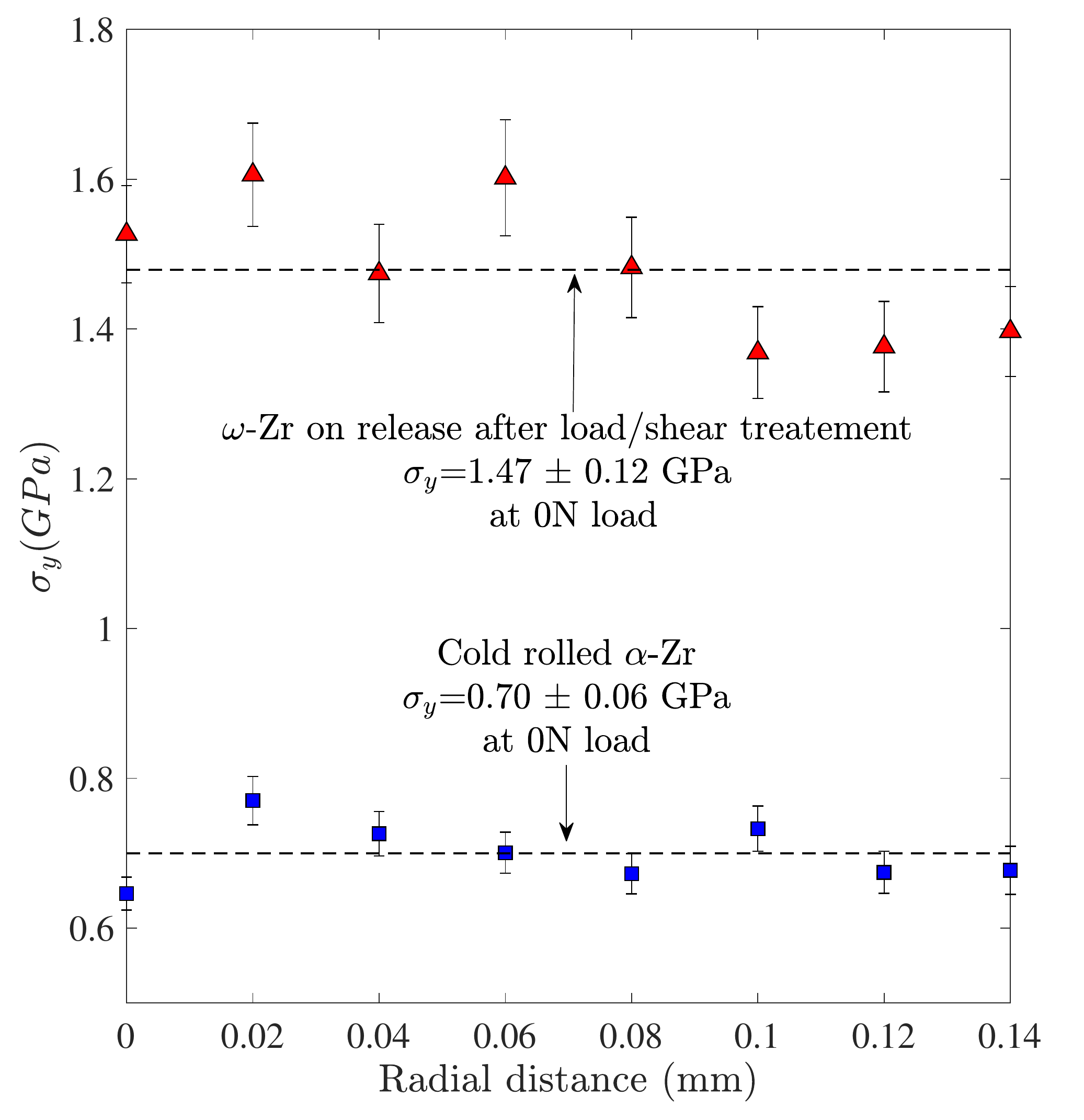}
\caption{Yield strengths of  $\alpha$-Zr and $\omega$-Zr estimated using peak broadening method \cite{Zhao-APL-2007}.}
\label{fig:ysfrompb}
\end{centering}

\end{figure}

\begin{subsection} {Utilizing x-ray peak broadening} \label{broadening}

The yield strengths of $\alpha$ and $\omega$ phases were also estimated using the peak broadening method \cite{Zhao-APL-2007} near the center of  a sample as shown in Fig. \ref{fig:ysfrompb}. The obtained yield strength of $\alpha$-Zr and $\omega$-Zr are comparable to that from the hardness measurements.

It is worth mentioning here that Zhao et al. reported the yield strength of $\alpha$ and $\omega$ phases as 0.18 GPa and 1.18 GPa, respectively \cite{Zhao-APL-2007} (nearly 6 times higher for $\omega$-Zr than for $\alpha$-Zr).  While our $\sigma_y $ for $\omega$-Zr is relatively close to that in \cite{Zhao-APL-2007}, our yield strength for annealed $\alpha$-Zr is much larger than that in \cite{Zhao-APL-2007}. The reason for this is that our sample is subjected to large plastic deformation during compression, while Zhao et al.  performed experiments on an annealed Zr sample in multi-anvil press, where  nonhydrostatic stresses are small and hardly introduce any strain hardening. Using deviations in yield strength values from mean value at  all the scanning positions across culet diameter, the estimated error in yield strength for $\alpha$-Zr and $\omega$-Zr are $\pm 0.06$ GPa and $\pm 0.12$ GPa, respectively using the x-ray peak broadening method.

\end{subsection}

\begin{subsection} {Utilizing pressure gradient method} \label{gradient}

The yield strength $\sg_y$ was also estimated  based on a simplified equilibrium equation combined with plastic friction and Tresca  plasticity criterion $dp/dr= -\sg_y/h$. We obtained the value of   $\sim$0.46 GPa for
$\alpha$-Zr (Fig. \ref{fig:ysfrompd}). The pressure distribution in the mixture does  not exhibit any visible signatures of PT such as, plateaus or change in slope \cite{Blank-Estrin-2014,Levitas-Zarechnyy-PRB-RDAC-10,Feng-Levitas-MSEA-16,Feng-Levitas-Mehdi-MSEA-18}. It is practically linear in the major part of a sample, like in simulations without PT or with PT and equal yield strength of phases \cite{Levitas-Zarechnyy-PRB-RDAC-10}. The estimated value of  $\sigma_y $ in the mixture grows from $\sim$0.46 GPa   to 0.93 GPa for $\omega$-Zr under further deformation. Since these values are smaller than those obtained above by two different methods, we may conclude that the shear stress did not reach the yield strength in shear at the contact surface and contact friction should obey the Coulomb law  as the only currently existing alternative
\cite{Feng-Levitas-MSEA-16,Feng-Levitas-Mehdi-MSEA-18,Levitas-Zarechnyy-PRB-RDAC-10,Levitas-etal-NPJ-CM-19}. This also shows the issues with determining of the yield strength based on the pressure gradient when we cannot prove that the friction stress has reached the yield strength in shear.

\begin{figure}[h!]
\begin{centering}
\includegraphics[width=0.8\linewidth]{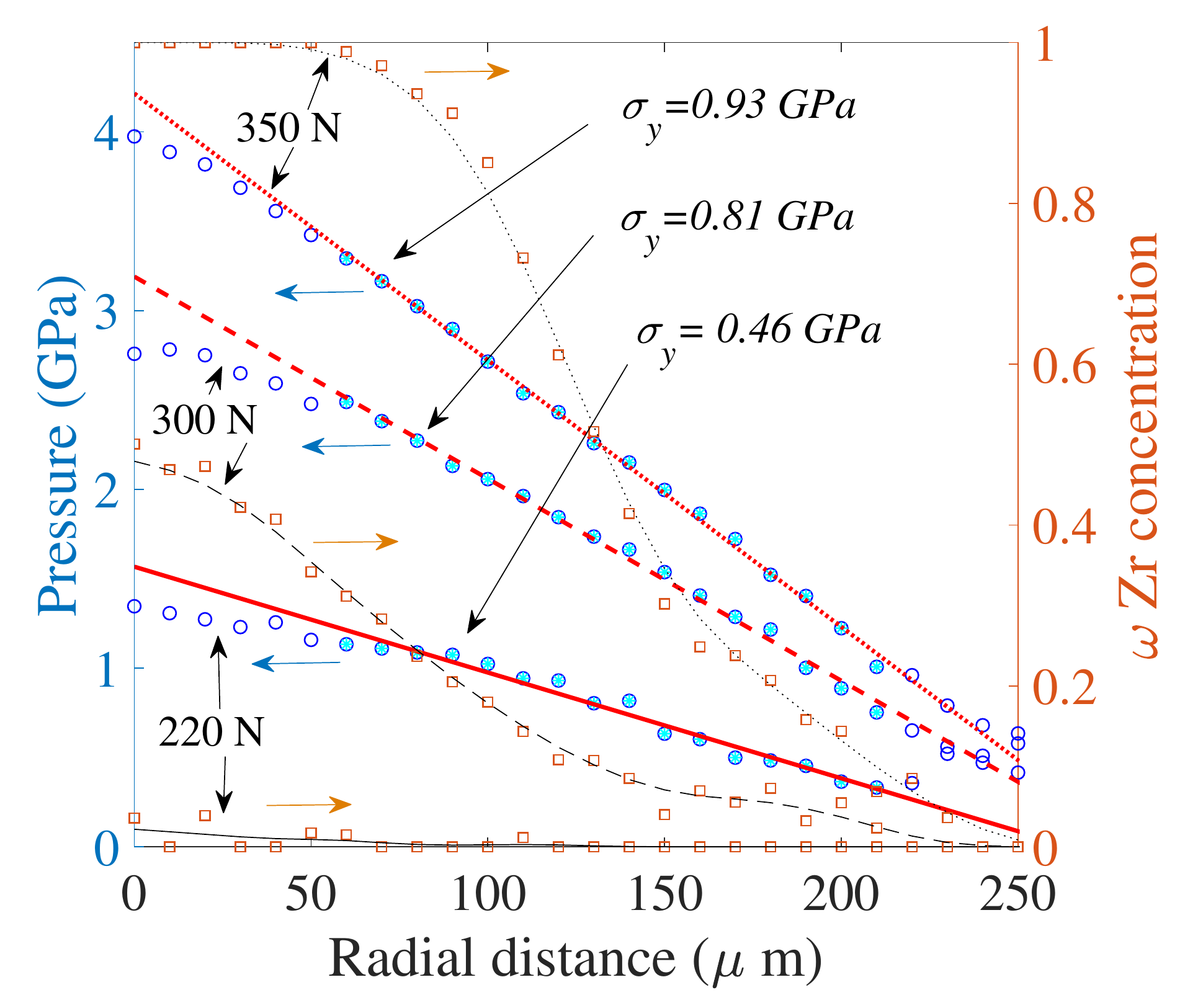}
\caption{Yield strengths of Zr estimated using the pressure gradient method. Here pressure distribution in mixture was used to estimate pressure gradient. Left y-axis corresponds to pressure whereas right y-axis corresponds to concentration of $\omega$-Zr phase.  }
\label{fig:ysfrompd}
\end{centering}

\end{figure}

\end{subsection}
\end{section}
}
{\color{black}
\begin{section}{Evaluation of accumulative plastic strain $q$}
Accumulated plastic strain averaged over the sample thickness has been determined at the sample center, where the strain state is a unidirectional compression. Let us define the following sample thicknesses:
initial and current thicknesses $h_0$ and $h$; thickness $h_e$ after elastic unloading from the current thickness, which occur
without reverse PT;  thickness $h_p$ at stress-free state after complete reverse PT to  $\alpha$ phase. Reverse PT can occur during heating or in thought experiment, as it is usually assumed in the large-strain theory for coupled elastoplastic deformations and PTs \cite{Feng-Levitas-IJP-BN-DAC-17}.
Since the ratio of lengths represents the corresponding deformation gradient, we introduce the multiplicative decomposition of the deformation gradient in the thickness direction $F$ into elastic $F_e$, transformational $F_t$, and plastic $F_p$ contributions:
\bey
F:= \frac{h}{h_0}=F_e F_t F_p; \qquad   F_e:= \frac{h}{h_e}; \quad   F_t:=\frac{h_e}{h_p};  \quad F_p=\frac{h_p}{h_0},
 \label{mult}
\eey
where $:=$ means equal by definition. Multiplicative decomposition Eq. (\ref{mult})  results in additive decomposition of logarithmic strains
\bey
ln F=ln \frac{h}{h_0}=ln F_e + ln F_t + ln F_p  \Rightarrow  ln \frac{h_0}{h}=ln \frac{h_e}{h} + ln\frac{h_p}{h_e} + ln \frac{h_0}{h_p}.
 \label{lnmult}
\eey
The accumulative plastic strain $q$ for uniaxial compression is $q=ln \frac{h_0}{h_p}$ and our goal is to show that elastic and transformation logarithmic strains are negligible in comparison to total or plastic strains.
The main quantitative results of the current paper related to determination of PT kinetics were obtained in the  pressure range from 1.2 to 4.0 GPa, so we will make estimates for these two pressures only.

{\it Total strain.} For $p_1=1.2 $GPa we have $h_1=70\, \mu m$
and for $p_2=4 $GPa the thickness is $h_2=50\, \mu m$. With $h_0=135\, \mu m$, we obtain $ln \frac{h_0}{h_1}=0.657$ and $ln \frac{h_0}{h_2}=0.993$.

{\it Transformation strain.} Based on Table S2, the maximum negative logarithmic volumetric transformations strain for complete PT is $ln(23.272/22.870)= 0.017$. If transformational compression occurs isotropically, then $ln\frac{h_p}{h_e}=0.017/3=0.006$. To obtain the upper bound, which is essential to overestimate, we assume that the transformation  strain occurs uniaxially with zero lateral strains, then $ln\frac{h_p}{h_e}=0.017$.

{\it Elastic strain.} To evaluate elastic strain, we will use the Hooke's law:
\bey
\vep_{ez}:= h_e/h-1= (\sg_z-2 \nu \sg_r)/E,
 \label{hooke}
\eey
where $\vep_{ez}$ is the axial elastic strain, $\sg_z$ and $\sg_r$ are the axial and radial stresses, and $E$ and $\nu$ are the Young modulus and the Poisson ratio. At zero pressure, the bulk moduli  $K_{\alpha}=92.2$ GPa and $K_{\omega}=102.4$ GPa (see Table S2),  the shear moduli $G_{\alpha}=36.3$ GPa and  $G_{\omega}=45.1$ GPa \cite{Liu-Zhao-JAP-08}, where from
$\nu_{\alpha}=0.326$ and $\nu_{\omega}=0.308$,    $E_{\alpha}=96.3K$ GPa and $E_{\omega}=118.0$ GPa,
in good correspondence with \cite{Liu-Zhao-JAP-08}. Both $E$ and $\nu$ increase with the pressure \cite{wangetal-2011}; neglecting this dependence for simplicity, we overestimate elastic strain.

Using $p=(\sg_z+2 \sg_r)/3$ and $\sg_z- \sg_r=\sg_y$ as well as $\sg_y^{\alpha}=0.7$  GPa and $\sg_y^{\omega}= 1.47$ GPa respectively, we obtain that for $p=1.2$ GPa in  $\alpha-$Zr $\vep_{ez}=0.0108$ and $ln (h_e/h)=0.0107$,  and for $p=4$ GPa in $\omega-$Zr $\vep_{ez}=0.0239$ and $ln (h_e/h)=0.0236$.
Thus, for $p=1.2$ GPa in  $\alpha-$Zr, $ln (h_e/h)=0.0107$ is just 1.6\% of the total strain $ln \frac{h_0}{h_1}=0.657$, and
for $p=4$ GPa in  $\omega-$Zr, $ln (h_e/h)+ln (h_p/h_e)=0.041$ is just 4.1\% of the total strain $ln \frac{h_0}{h_2}=0.993$.

That means that with the error less than 4.1\%  in the pressure range up to 4 GPa,  elastic and transformational strain can be neglected in comparison with the plastic strain and $q$ can be estimated as $ln(h_0/h)$, which is used in the main text.
 \end{section}
}

\begin{section}{Thickness estimation}
For sample thickness estimation, 2D/1D x-ray absorption scans were recorded. X-ray absorption was estimated  using the current measured at the reference ionization chamber, ($I_\circ$) (before sample), and at photo diode at beam stop, ($I$) (after sample). These current values were corrected for absorption in diamond anvils.

\begin{figure}[h!]
\vspace{-3mm}
\begin{centering}
\includegraphics[width=0.6\linewidth]{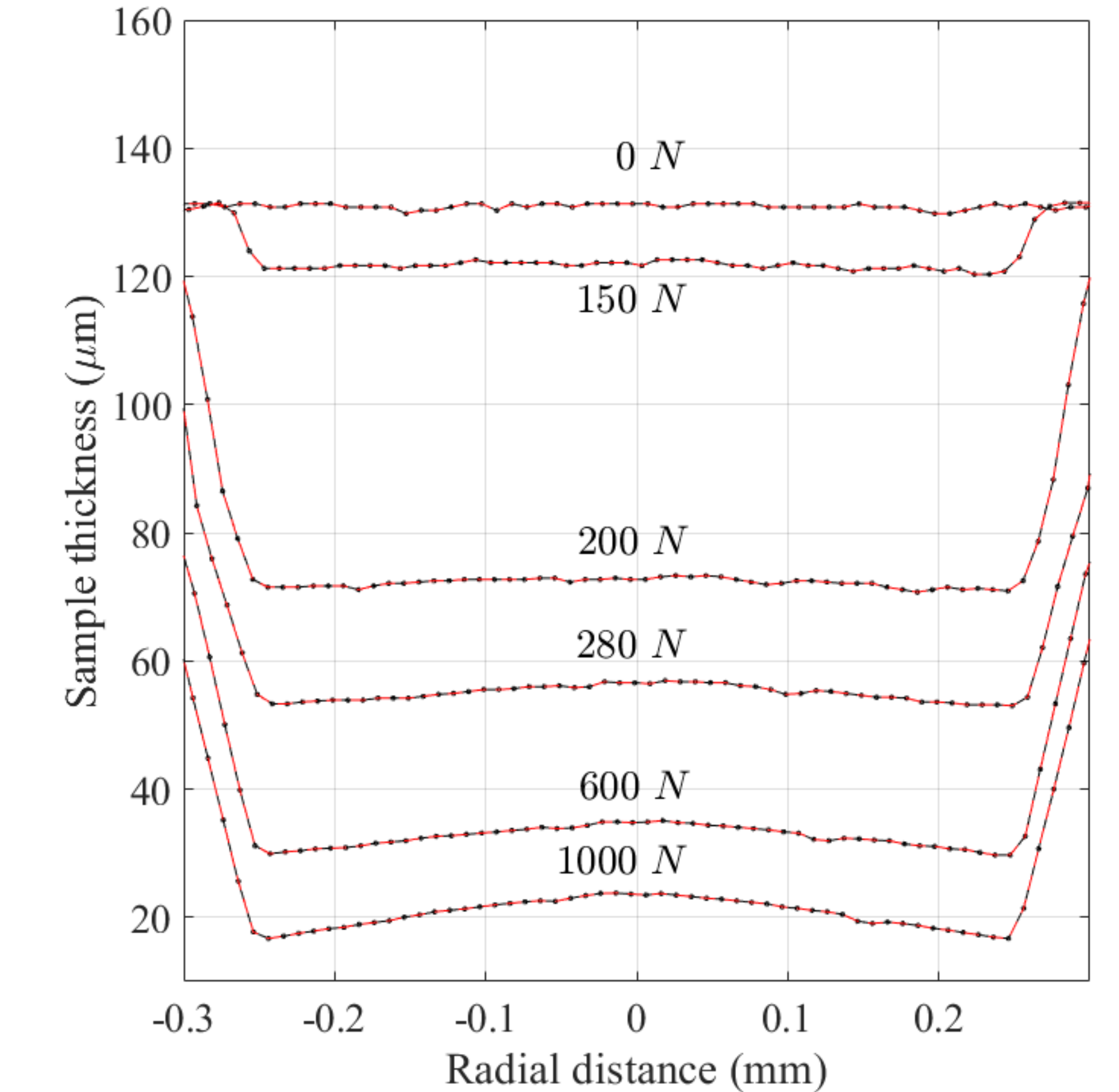}
\caption{Thickness profile of Zr sample at various loads estimated using absorption law.}
\label{fig:thicknessprofile}
\end{centering}
\end{figure}

Subsequently, thickness at each scanning position was estimated using the absorption law equation,
\bey
I/I_\circ=exp(-\mu \rho h),
 \label{int}
\eey
like in \cite{Li-etal-PNAS-17}. Here $I$, $I_\circ$, $\mu$, $\rho$ and $h$ represent the
X-ray intensity after the sample, incident X-ray intensity before the sample, mass attenuation coefficient, mass density, and thickness of the sample. For correct thickness estimation at each scanning position, density was calculated using lattice parameters of Zr at each position. Fig. \ref{fig:thicknessprofile} shows thickness profile of sample across culet at a few representative loads.

The error in thickness measurements using absorption method can be attributed to noise in the currents measured from ionization chamber/photo diode detector, error in density estimation and error in mass attenuation coefficients, $\mu$. {\color{black}The mass attenuation coefficients, $\mu$, used for the thickness estimations of Zr sample are 24.85 $cm^2/g$ and 11.39 $cm^2/g$ for 30 keV X-rays (at beamline 16-ID-B) and 40 keV X-rays (at beamline 16-BM-D)  respectively \cite{Hubbel-2004}.

Assuming accuracy of $\mu$ is equal to the precision of reported values, the error in $\mu$ are of the order of $\sim$ 0.1\%. Error in the current measurements as estimated from the noise is of the order of  0.3\%. The other contribution to error is due to error in the mass density estimation which depends on the error in pressure estimation. As error in our pressure estimation is $ \pm 0.2 GPa$, the estimated change in density due to this is well within $\sim$ 1\% in our case.

Under compression, density can be estimated as
\bey
\frac{1}{\rho}=\frac{1}{\rho_o}\frac{V}{V_o},
 \label{density}
\eey
where $\rho_o$ and $V_o$ are the mass density and unit cell volume at ambient pressure and $V$ is the unit cell volume at given pressure.
The main quantitative results of the current paper related to determination of PT kinetics were obtained in a pressure range from 1.2 to 4.0 GPa (see loading path in Fig. 2 of the main text). For 4 GPa, $V/ V_o$ is 0.97  for $\alpha$ -Zr. Thus, even if we  use ambient density $\rho_o$ in Eq. (\ref{int}), our error will not exceed 3\%.

Convoluting all these errors, estimated error in thickness, $\Delta h$ is maximum up to $ \pm 1  \mu m$.
A similar error in $\Delta h$ is obtained in selected measurements of the sample thickness after complete load release,
corrected by elastic strain.
Accumulated plastic strain, $q$, has been calculated as $ln(h_\circ/h)$, where $h_\circ$ and $h$ are the initial and final thickness of sample at symmetry axis. Hence, using this method of error estimation of function of measured parameters with the same source of error, the  error in $q$ has been estimated as $\Delta q=2\Delta h/h$ at each data point.
}

\end{section}